%% file: main.tex
\documentclass[letterpaper,twocolumn,10pt]{article}

\newif\ifSPACEHACK
\SPACEHACKfalse

\newif\ifDEBUG
\DEBUGtrue
\DEBUGfalse

\newif\ifANONYMOUS
\ANONYMOUStrue
\ANONYMOUSfalse

\newif\ifEXTENDED
 \EXTENDEDfalse
\newif\ifARXIV
\ARXIVtrue
\input{misc/typesetting}

\usepackage{usenix}
\usepackage{multicol}
\usepackage{multirow}

\usepackage{tikz}
\usepackage{amsmath}

\usepackage{filecontents}

\usepackage{tablefootnote}

\usepackage{cite}
\usepackage{amsmath,amssymb,amsfonts}
\usepackage{algorithmic}
\usepackage{graphicx}
\usepackage{textcomp}
\usepackage{placeins}
\usepackage{xcolor}
\ifARXIV
\else
\def\BibTeX{{\rm B\kern-.05em{\sc i\kern-.025em b}\kern-.08em
    T\kern-.1667em\lower.7ex\hbox{E}\kern-.125emX}}
\usepackage[available]{usenixbadges}
\fi
\begin{document}
%-------------------------------------------------------------------------------

% % make title bold and 14 pt font (Latex default is non-bold, 16 pt)
% \title{\Large \bf Formatting Submissions for USENIX Security 2026:\\
%   An (Incomplete) Example}

\input{data/data}

%don't want date printed
\date{}


% \title{\Large \bf Artifact Signing in Five Public Software Package Registries: \\ Adoption Measurements and Influencing Factors}

\newcommand{\MyTitle}{}

\renewcommand{\MyTitle}{Exploring the Implementation of Software Supply Chain Security Methods in Mitigating Risks: A Case Study of Software Signing}
\renewcommand{\MyTitle}{An Industry Interview Study of Software Signing for Supply Chain Security}
\renewcommand{\MyTitle}{Bad Workman or Faulty Tools? Tooling as a Factor of Software Signing Adoption: The Case of Sigstore}
\renewcommand{\MyTitle}{A Bad Workman or Faulty Tools? Examining Tooling as a Factor in Software Signing Adoption: The Case of Sigstore}
\renewcommand{\MyTitle}{A Bad Workman or Faulty Tools? Examining Tooling as a Factor in Software Signing Adoption in the Sigstore Ecosystem}
\renewcommand{\MyTitle}{Why Johnny Signs with Sigstore: Examining Tooling as a Factor in Software Signing Adoption in the Sigstore Ecosystem}
\renewcommand{\MyTitle}{Adoption Dynamics of identity-based Signing Tools: A Practitioner-Focused Usability Study of Sigstore}
\renewcommand{\MyTitle}{Examination of identity-based Signing Adoption: A Usability Case Study of Sigstore}
\renewcommand{\MyTitle}{Why Johnny Signs with Identity-Based Tools: \\ A Usability Case Study of Sigstore}
\renewcommand{\MyTitle}{Why Johnny uses Identity-Based Software Signing Tools: \\ A Usability Case Study of Sigstore}
\renewcommand{\MyTitle}{Why Johnny Adopts Identity-Based Software Signing Tools: \\ A Usability Case Study of Sigstore}
\renewcommand{\MyTitle}{Why Johnny Adopts Identity-Based Software Signing: \\ A Usability Case Study of Sigstore}

\title{\MyTitle}

\ifANONYMOUS
    \author{Anonymous author(s)}
    
\else
    \author{
    {\rm Kelechi G.\ Kalu}\\
    Purdue University \\ 
    kalu@purdue.edu
    \and
    {\rm Sofia Okorafor}\\
    Purdue University \\ 
    sokorafo@purdue.edu
     \and
    {\rm Tanmay Singla}\\
    Purdue University \\ 
    singlat@purdue.edu
     \and
    {\rm Sophie Chen}\\
    Carnegie Mellon University \\ 
    scchen@andrew.cmu.edu % Hack to get Santiago to be on line 2
     \and
    {\rm Santiago Torres-Arias}\\
    Purdue University \\ 
    santiagotorres@purdue.edu
     \and
    {\rm James C.\ Davis}\\
    Purdue University \\
    davisjam@purdue.edu 
    % copy the following lines to add more authors
    % \and
    % {\rm Name}\\
    %Name Institution
    } % end author
\fi

\maketitle
% \thispagestyle{empty}

%-------------------------------------------------------------------------------

\begin{abstract}
Software signing is the most robust method for ensuring the integrity and authenticity of components in a software supply chain.
Legacy key-managed signing tools (\eg OpenPGP) burdened practitioners with key management and signer identification, creating both usability challenges and security risks.
A new class of identity-based signing tools automate many of these concerns, but little is known about their usability and its effect on their adoption and effectiveness in practice. A usability evaluation can clarify the extent to which identity-based designs succeed and highlight priorities for improvement.

To fill this gap, we conducted the first usability study of Sigstore, a pioneering and widely adopted exemplar of identity-based signing. 
Through interviews with 17 industry experts, we examined (1) the problems and advantages associated with practitioners’ tooling choices, (2) how and why their signing-tool usage has evolved over time, and (3) the contexts that cause usability concerns.
Our findings illuminate the usability factors of identity-based signing tools and yield recommendations for toolmakers, adopting organizations, and the research community. Notably, components of identity-based tooling exhibit different levels of maturity and readiness for adoption, and integration flexibility is a common pain point but potentially mitigable through plugins and APIs.
Our results will help identity-based signing toolmakers further strengthen software supply chain security.
\end{abstract}


% \begin{IEEEkeywords}
% Software Signing, Software supply chain
% \end{IEEEkeywords}

% \fi
%\IEEEpeerreviewmaketitle

% \KC{First sentence of second  paragraph should come in the first paragraph, and second paragraph should be about signing, remove excess detail in the second paragraph about the previous work.}
% \KC{3rd paragraph: A good topic sentence and itemize the study points of the work. Remove excess detail about subjects}

\section{Introduction}
\label{sec:intro}

The reuse of software components in modern software production creates complex supply chains that are susceptible to the unauthorized introduction of code~\cite{willett2023lessons, OSSRA_2024, benthall_assessing_2017,ladisa_sok_2023}. 
%These include unauthorized code injections, package confusion, and the distribution of malicious package versions~\cite{ladisa_sok_2023}.
To mitigate this risk, engineers need \textit{provenance}—evidence of actor authenticity and artifact integrity~\cite{okafor_sok}.
%—through \textit{software signing}~\cite{okafor_sok, schorlemmer2025establishing}.
Software signing provides the strongest guarantee of provenance currently available~\cite{Schorlemmer_Kalu_Chigges_Ko_Ishgair_Bagchi_Torres-Arias_Davis_2024} and has thus been widely advocated by academia, industry consortia, and government regulators~\cite{whitehouse_cybersecurity_2021, cooper_protecting_2018, cncf_2021, chandramouli2024strategies}.

Like other security practices, the effectiveness of software signing in practice depends on the availability and usability of tools that support it.  
Traditional signing tools such as OpenPGP are famously challenging to use~\cite{whitten1999johnny}, with several “Why Johnny Can’t Encrypt”–inspired studies consistently revealing usability challenges that limit adoption~\cite{whitten1999johnny,sheng2006johnny,ruoti2016johnny}.  
In recent years, identity-based systems (\eg Sigstore~\cite{newman_sigstore_2022}, OpenPubKey~\cite{heilman_openpubkey_2023}, SignServer~\cite{signserver_about_2025}) have been introduced that automate much of the process using short-lived keys and external identity providers~\cite{heilman_openpubkey_2023}.  
However, despite growing adoption driven by recent supply chain attacks and efforts to lower usability barriers~\cite{newman_sigstore_2022, schorlemmer2025establishing, merrill2023speranza}, we lack evidence about the usability of these identity-based tools and need to understand how that affects adoption in practice~\cite{lewis2014usability}. 
The innovations in identity-based approaches to signing motivate the need to evaluate how usability concerns affect their adoption and effectiveness. % in this new landscape.
A usability evaluation can clarify the extent to which identity-based designs succeed, highlight how they compare to legacy key-managed signing approaches, and surface priorities for improvement.

To this end, we offer the first empirical study of the usability of identity-based software signing tools.
Our work examines Sigstore, a pioneer identity-based signing tool that has seen rapid uptake across major open‐source ecosystems\cite{sonatype2024central, Hinds_Blauzvern_2023, mlarson2024python}, cloud‐native projects, and corporations\cite{Vaughan-Nichols_2022_b, Sigstore_2022a, Sigstore_2022b, Sigstore_2022c}, involving a large number of engineers worldwide. 
%Focusing on the exemplar tool Sigstore,
We interviewed 17 experienced security practitioners to investigate the usability concerns that drive or discourage \sigs adoption. 
We adopted an exploratory data analysis approach, because this study addresses a new class of tools that has not previously been analyzed. 
%Specifically, we started with thematic analysis, inductively drawing patterns from practitioners’ narratives, and then situated these findings within a formative usability framework~\cite{cresswell2020developing}. 
%This combination allowed us to capture the breadth of practitioner perspectives and to ground our study in the context of prior formative usability research.
Our analysis centered on two aspects: practitioners’ direct experiences with Sigstore and their perceptions of its usability relative to other signing tools.

Our study highlights the usability factors practitioners weigh when adopting identity-based software signing tools.
%We articulate the challenges practitioners face and the perceived merits of adoption.
We show that while identity-based signing tools ease legacy key-managed challenges, adoption is still constrained by integration hurdles, privacy concerns, and organizational factors. Second, our results provide adopting organizations with insight into the readiness levels of these tools, helping them manage adoption risks. Finally, we offer toolmakers and designers concrete directions for improving identity-based signing tools. Together, these findings provide formative feedback situated in real-world usage contexts. %that clarifies the drivers of real-world usage and the contexts in which they matter.

%They also corroborate literature on IT tool adoption, showing that many usability factors recur across different software systems.

% In summary, we make the following contributions:
% \noindent
% \begin{enumerate}
% 	\item We report on the difficulties, and advantages of using \sig (a identity-based \signs tool) (\cref{sec: sigstore_strength}, \cref{sec: sigstore_weakness}).
%     \item We report on how practitioners \sign tools evolve (\cref{sec:RQ3}).
%     \item We also identified the factors that informs the decisions of practitioners to switch tooling options (\cref{implem_challenges}).
% \end{enumerate}
In summary, we make the following contributions:

\begin{enumerate}
  \item We report on the difficulties and advantages of using \sig, an identity-based \signs tool. 
  % \item We discuss how practitioners' choice of \signs tools evolve over time \JD{the tools themselves, or the tools they select? Also, not sure about `evolve', it implies a continual process, and I guess we see discrete state changes?} and the factors that influence practitioners' decisions to switch tooling options.
  \item We discuss how practitioners’ choices of software signing tools change over time and the factors influencing those decisions.
  
  \item We discuss how organizational deployment contexts influence the usability concerns raised by practitioners.
\end{enumerate}

% \ifARXIV
% \myparagraph{Significance of Study}
% \else
% \myparagraph{Significance to (Automated) Software Engineering}
% \fi

\noindent\myparagraph{Significance}
% Our work offers a number of importance to the securoty of the software supply chains;
% First, By unpacking the usability factors that drive or hinder Sigstore’s adoption, we not only inform the ongoing refinement of its workflows but also provide a template for evaluating and improving similar identity-based (identity-based) signing solutions.
% Also, since Sigstore exemplifies a broader class of tools that embed security into automated build and deployment processes, our findings can be potentially generalized to guide the design and implementation of other automated software‐supply‐chain technologies, ultimately advancing the usability—and thus the security—of automated software engineering practices.
% Second, the wide usage of sigstore (adopted by kubernetes, maven, pypi etc),means that a formative usability analysis of sigstore would signifi cantly offer an improvement template which would  improve the effectiveness and quality of the use of this tool to provide more securely guaranteed signatures.  
%
%
Our work contributes to the security of software signing tools and, more broadly, to software supply chains.
By unpacking the usability factors that promote and hinder \sigs adoption, we inform the ongoing refinement of its workflows and provide a template for evaluating and improving similar identity-based signing solutions.
Because \sig exemplifies other identity-based software signing tools, our findings can be generalized to guide the design and implementation of this class of software, ultimately advancing usability (and thus security) across the software supply chain.

% \JD{I cut the rest of this, felt too subtle}
%\textit{Second}, by evaluating the context of use across different organizations, we pinpoint the usability and organizational frictions that hinder deployment of Sigstore and other identity-based signing tools. Addressing these frictions increases the coverage of verifiably signed artifacts and attestations, shrinking the attack surface across pipelines.

\section{Background \& Related Works}
\label{sec:Background}

\iffalse
In this section, we discuss the workflow of software signing and the distinction of identity-based tools in contrast with legacy key-managed tools (\cref{sec:Background-SoftwareSigning}),
  %software signing tools (\cref{sec:Background-SoftwareSigning_tools}),   relevant empirical studies (\cref{sec:Background-SoftwareSigning_research}), 
  usability studies in software signing (\cref{sec:Background-Usability}),
  and
  the Sigstore ecosystem where we situate our work (\cref{sec:Method-CaseStudyContext}).
  \fi
  
In this section, we review
  software signing technologies (\cref{sec:Background-SoftwareSigning})
  and
  usability assessments (\cref{sec:Background-Usability}).

\subsection{Software Signing}
\label{sec:Background-SoftwareSigning}

Software signing ensures the authorship and integrity of a software artifact by linking a maintainer's cryptographic signature to their software artifact.
%Fundamentally, software signing is based on the public-key cryptography model ~\cite{Schorlemmer_Kalu_Chigges_Ko_Ishgair_Bagchi_Torres-Arias_Davis_2024}, and a legacy key-managed signing workflow follows a three-stage process of creation, distribution, and verification. 
From literature, the implementation of software signing is shaped by several factors:
  the software artifacts to be signed (\eg driver signing\cite{cooper_security_2018}),
  the root-of-trust design (\eg public key infrastructure/PKI~\cite{zimmer2016establishing}),
  and
  ecosystem policies (\eg PyPi\cite{Kuppusamy2014PEP} and Maven\cite{MavenWorking}).
\subsubsection{Software Signing Tools}

Software signing tools can be broadly grouped into two families~\cite{schorlemmer2025establishing}. 
\emph{Legacy key-managed} tools such as GPG~\cite{GNUPG_2024} follow a conventional public-key 
model, relying on long-lived asymmetric key pairs and delegating key-management responsibilities 
to the signer; trust is typically established through endorsements by other signers. In contrast, 
\emph{identity-based} signing tools (described as “Next-generation signing” by Schorlemmer 
\etal~\cite{schorlemmer2025establishing}) automate key management by issuing short-lived, 
ephemeral certificates tied to a signer’s identity via external OpenID Connect (OIDC) or OAuth 
providers. This “keyless signing” approach, exemplified by Sigstore~\cite{newman_sigstore_2022}, 
OpenPubKey~\cite{heilman_openpubkey_2023}, and SignServer~\cite{signserver_about_2025}, reduces long-term key 
handling and often embeds transparency and auditability through append-only logs. We discuss the 
architecture and workflow of identity-based signing in~\cref{sec:Method-CaseStudyContext} using Sigstore as a case study.

% \subsubsection{Effects of Signing Tools on Software Signing Adoption}
\subsubsection{Tooling Effects on Signing Adoption}
\label{sec: bg-effects_signing_tools}

Previous studies~\cite{merrill2023speranza, Schorlemmer_Kalu_Chigges_Ko_Ishgair_Bagchi_Torres-Arias_Davis_2024, usenix_2025_signing_interview_kalu} recognize that tooling influences whether and how software signing is adopted. 
For example, Schorlemmer \etal~\cite{Schorlemmer_Kalu_Chigges_Ko_Ishgair_Bagchi_Torres-Arias_Davis_2024}  perform a measurement study of several software package registries, 
identifying tooling as a key determinant of signature quality; their results indicate that “dedicated tools” 
lead to higher-quality software signatures. Kalu \etal~\cite{usenix_2025_signing_interview_kalu} examine the implementation and adoption of signing 
in real-world organizational settings, highlighting tool usability as a central technical challenge for using 
software signing in practice. Their study focuses on how organizations incorporate signing into their existing 
software delivery processes, without centering on any particular tool. 
Although these works highlight tooling, and, in the case of Kalu \etal, usability, as important concerns, they do so only at a high level and provide little concrete guidance on what makes a signing tool usable or unusable in practice.

In this study, by contrast, we examine the specific usability of identity-based signing tooling, using Sigstore as a case study. We concretize formative usability concerns for Sigstore and extend Kalu \etals~\cite{usenix_2025_signing_interview_kalu} work by specifying which features and design choices make identity-based signing tools like Sigstore usable or unusable in practice, and by translating these findings into concrete improvement guidance.

\subsection{Usability Studies in Software Signing}
\label{sec:Background-Usability}

\subsubsection{Usability --- Definition and Evaluation Approaches}

Tools enable engineers to adopt new software development practices~\cite{Bruckhaus_Madhavii_Janssen_Henshaw_1996}. A usable tool is one whose functionality effectively supports its intended purpose, achieving product goals~\cite{lewis2014usability, sasse2005usable}. Studies assessing implementation strategies of tools are typically framed as usability evaluations~\cite{lyon2021cognitive}.

\begin{figure*}
    \centering
    \includegraphics[width=0.85\textwidth]{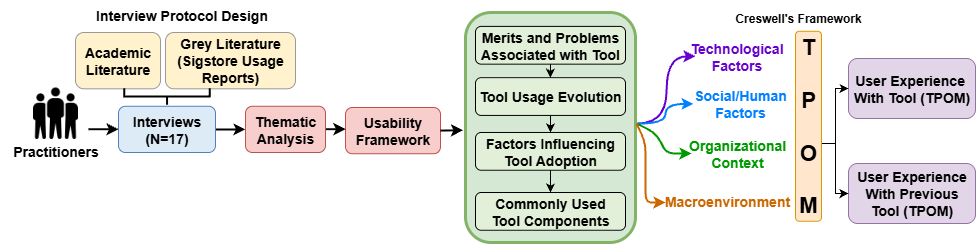}
    \vspace{5pt}
    \caption{
    Study Methodology and Usability Framework. 
    We developed our interview protocol by reviewing academic and grey literature on software signing, then conducted 17 practitioner interviews. 
    Data were analyzed through thematic analysis and mapped onto Cresswell’s formative usability evaluation framework~\cite{cresswell2020developing}.
    The framework organizes usability concerns into four factors: Technological (T) — technical characteristics such as performance, ease of use, and data integrity; Social/Human (P) — user experience, and correct use of features; Organizational (O) — internal factors such as training, leadership, and resource support; and Macroenvironmental (M) — external influences including regulation, economics, and ecosystem community.
    }
    % \vspace{3pt}
    \label{fig:methodology}
\end{figure*}

To evaluate a tool’s usability, studies typically follow a \emph{usability evaluation framework}~\cite{lyon2021cognitive}. There are two kinds of approaches. %\footnote{In the usability literature, \textit{summative usability} is sometimes referred to simply as \textit{usability}~\cite{LewisSauro2021UsabilityUX}, while \textit{formative usability} is also described as \textit{formative evaluations} or \textit{formative studies}~\cite{Rohn2002UsabilityPractice}.}
Summative work measures a tool’s effectiveness~\cite{iso9241-11, rusu2015user}, \eg by task success rates, completion time, and user satisfaction.
Formative assessments provide feedback to guide ongoing improvement~\cite{lewis2014usability, theofanos_usability}.
Given our aims, we took a \textbf{formative} approach. We considered several formative frameworks~\cite{chan2022practical, shah2023my, andre2000testing, strifler2024development} and found that Cresswell \etals four-factor framework~\cite{cresswell2020developing} best explained our data.
That framework is summarized in~\cref{fig:methodology}.
Usability depends on the context in which a tool is used.
Many factors, such as organizational objectives, personnel, policies, and compliance requirements, can make a tool usable in one setting and impractical in another~\cite{whitten1999johnny, lennartsson2021exploring}.
This contextual aspect informs our study design. %incorporated into usability evaluation. %assessing tool usability. %when designing and evaluating to what extent and in what ways a tool is usable.
%Thus, because usability shapes adoption and adoption shapes security, evaluating security tools such as identity-based software signing tools in context-aware manner is essential to understand and improve the real-world impact of security research.

% These results underscore that systematic, context-aware usability evaluation is not merely diagnostic; it is a practical lever for improving adoption and, by extension, security outcomes. 

\subsubsection{Empirical Studies of Signing Usability}
\label{sec: bg_empirical_research}
% \myparagraph{Signing Tooling Usability Studies}
% \JD{Not sure why there is a myparagraph here?}
\iffalse
The usability of legacy key-managed (digital) signing tools has been widely studied, though often within the context of communication encryption rather than broader software signing applications. Whitten and Tygar’s seminal study, \textit{Why Johnny Can't Encrypt}~\cite{whitten1999johnny}, found that PGP 5.0’s user interface was too complex for non-expert users, making encryption and signing inaccessible. Sheng et al.~\cite{sheng2006johnny} later observed that PGP 9’s interface had not significantly improved, particularly in key verification, encryption transparency, and digital signing usability. Other studies~\cite{braz_security_2006, ruoti_confused_2013, reuter_secure_2020} have highlighted persistent usability issues, including the complexity of the public-key model.

Further research has explored the differences between manual and automated signing tools. Fahl et al.~\cite{fahl_helping_2012} and Ruoti et al.~\cite{ruoti_private_2015} found no significant usability differences between these approaches. Meanwhile, Atwater et al.~\cite{atwater_leading_2015} demonstrated that users preferred automated solutions based on the PGP standard. While these studies provide insights into encryption tools, their relevance to broader software signing workflows is limited.
\fi
The usability of legacy key-managed signing tools, often studied in the context of encryption, remains a significant challenge. Whitten and Tygar’s seminal study, ``\textit{Why Johnny Can't Encrypt''}~\cite{whitten1999johnny}, found PGP 5.0’s interface too complex for non-experts, a problem that persisted in PGP 9~\cite{sheng2006johnny}, especially around key verification, transparency, and signing.
Subsequent work~\cite{braz_security_2006, ruoti_confused_2013, reuter_secure_2020} has underscored the public-key model’s inherent complexity.
Several works have subsequently examined the use and automation of signing in specific domains, such as social media conversations~\cite{fahl_helping_2012,ruoti_private_2015,atwater_leading_2015}.
% Most pertinent to software signing, a recent measurement study by Schorlemmer \etal offers a summative assessment --- that legacy key-managed software signing tools are unusable~\cite{ Schorlemmer_Kalu_Chigges_Ko_Ishgair_Bagchi_Torres-Arias_Davis_2024}.

\iffalse
Research comparing manual versus automated signing tools has yielded mixed results:
some works observed no clear usability advantage~\cite{fahl_helping_2012,ruoti_private_2015}, whereas Atwater \etal showed a user preference for automated PGP-based solutions~\cite{atwater_leading_2015}. 
\SC{Is not clear what manual vs automated signing tool means--would like example of manual tool (since automated tool example of PGP was given)}
\fi

These prior usability studies of signing tools are limited in applicability to software signing. Much of these works have focused on communication contexts such as email and message encryption, where users have less technical expertise~\cite{kjeldskov2005does,macdorman2011improved} than software engineers securing their software supply chains.
These studies also examine only legacy key-managed signing technologies.
We contribute the first usability study of identity-based software signing tools.

\section{\textbf{Knowledge Gaps \& Research Questions}}
\label{sec: knowledge_gaps}
Our literature review shows a gap in formative assessments of software signing tools. %identity-based software signing tools.
%studying signing tools within the context of software signing and software supply chain security.
Furthermore, no study has examined identity-based signing tools, which are structurally and functionally different from legacy key-managed signing tools. %, there remains a lack of usability evaluation in this context. 
We therefore ask:

\begin{itemize}[leftmargin=33pt, rightmargin=5pt, itemsep=0pt, parsep=0pt]
  \item[\textbf{RQ1:}] What usability strengths and weaknesses of identity-based software signing tools shape their adoption in practice?
  %\item[\textbf{RQ2:}] What factors influence practitioners’ decisions to switch from other signing tools to Sigstore over time?
  \item[\textbf{RQ2:}] What factors drive practitioners to adopt, retain, or replace software signing tools over time?
  % What factors drive practitioners to adopt, retain, or replace software signing tools over time?
  % What factors influence practitioners’ decisions to change the software signing tools they use?
\end{itemize}
% We choose a case study method (discussed next), so these RQs are focused on Sigstore.

\section{Methodology}
\label{sec:Method}

We investigated our research questions with a case study.
\cref{fig:methodology} summarizes our methodology. 
This section details our research design, including the study rationale~(\cref{sec:Method-Rationale}), case study context~(\cref{sec:Method-CaseStudyContext}), instrument design~(\cref{sec:Method-InstrumentDesign}), data collection~(\cref{sec:Method-DataCollection}), and analysis procedures~(\cref{sec:Method-DataAnalysis}).

%The Sigstore ecosystem consists of \signs tools designed and managed by the Linux Foundation. 
% Our study involved two phases: a semi-structured interview study and a survey.
% The interviews were conducted as part of Kalu \etal's study~\cite{}. 
% These interviews provided a holistic exploration of the various aspects of \signs tools and their influence on the adoption of \signs. To validate our interview findings, we conducted a survey targeting a broader group of Sigstore users.

Our Institutional Review Board (IRB) approved this work.
The protocol number is Purdue-IRB-2023-841.

%In the following sections, we discuss the justification for our use of a case-study approach ~\cref{sec:case_study}, instrument design ~\cref{sec:Method-InstrumentDesign}, data collection ~\cref{sec:Method-DataCollection}, data analysis ~\cref{sec:Method-DataAnalysis}, and limitations of our study ~\cref{sec:Threats}. 

% \subsection{Research Design and Rationale}
\subsection{Rationale for Case Study}
\label{sec:Method-Rationale}

\ST{Why deifine it here} \JD{Let's remove the preceding sentence and weave the citations there into the next sentence(s).} We selected the case study method for our research because it allows for an in-depth exploratory examination of our chosen case, enabling a deeper understanding of the usability factors that influence the adoption of a \signs tool in practice~\cite{Runeson_Höst_2009, yin_case_2018, Crowe_Cresswell_Robertson_Huby_Avery_Sheikh_2011, Gerring_2004}, rather than a broad but superficial cross-analysis~\cite{ralph_acm_2021}.
In selecting Sigstore, we adhered to the criteria outlined by Yin~\cite{yin_case_2018} and Crowe \etal~\cite{Crowe_Cresswell_Robertson_Huby_Avery_Sheikh_2011}. 
Specifically, we considered the intrinsic value and uniqueness of the case study unit (high), access to data (adequate), and risks of participation (low). 
In~\cref{sec:Discussion} we discuss generalizability.

\subsection{Case Study Context -- Sigstore}
\label{sec:Method-CaseStudyContext}
\iffalse
Due to the emergence of identity-based software signing and the growing adoption of Sigstore as a signing platform, we focus on the Sigstore tool specifically in our research.
\fi

\iffalse
Next, we provide a description of our selected case study-\emph{Sigstore}-Outlining its relative importance and prime position as  a state of the art identity-based software signing tool, its design and major components. 
\fi

% \JD{I would prefer not to cite ACM SIGSOFT directly, but rather one or more of the methods papers referenced therein.}
Following guidelines outlined by Runeson \& Höst~\cite{Runeson_Höst_2009}, we describe the importance of our selected case study context and the technical details of the system.

\subsubsection{Sigstore in Practice} 
% Sigstore is an exemplar of identity-based software signing tools. 
% It provides \emph{keyless signing}, where developers authenticate using existing OIDC identities (e.g., GitHub, Google) instead of generating and distributing long-lived keys. 
% It also issues short-lived (ephemeral) certificates that simplify key management and reduce long-term security risks compared to legacy key-managed signing tools. 
% Within 13 months of its public release in 2021, Sigstore recorded 46 million artifact signatures~\cite{Hinds_Blauzvern_2023}. 
% It has since gained broad adoption across open-source ecosystems (\eg PyPI~\cite{mlarson2024python}, Maven~\cite{sonatype2024central}, NPM~\cite{sigstore2024npmbeta}) and major companies (\eg Autodesk~\cite{Sigstore_2022c}, Verizon~\cite{Sigstore_2022a}, Yahoo~\cite{yahoo2024scaling}). 
% Given its rapid uptake, integration into critical ecosystems, and pioneering design features, Sigstore provides a compelling case study for examining the usability of identity-based signing tools.
Sigstore provides \emph{keyless signing}, where developers authenticate using existing OIDC identities (e.g., GitHub, Google) instead of generating and distributing long-lived keys. 
It issues short-lived (ephemeral) certificates that simplify key management and reduce long-term security risks compared to legacy key-managed signing tools. 
Within 13 months of its public release in 2021, Sigstore recorded 46 million artifact signatures~\cite{Hinds_Blauzvern_2023}. 
It has since gained broad adoption across open-source ecosystems (\eg PyPI~\cite{mlarson2024python}, Maven~\cite{sonatype2024central}, NPM~\cite{sigstore2024npmbeta}) and major companies (\eg Autodesk~\cite{Sigstore_2022c}, Verizon~\cite{Sigstore_2022a}, Yahoo~\cite{yahoo2024scaling}). 
Given its rapid uptake, integration into critical ecosystems, and pioneering design features, Sigstore provides a compelling case study for examining the usability of identity-based software signing tools.
% \SC{Let's do "Sigstore provides keyless, ephermal keys and siner authentication mech.... Within 13 months of its public release in 2021" instead--is easier to understand what sigstore does first as intro to paragraph}
% \SC{Also, am confused as to what "keyless, ephermal keys" means--this may warrant some elaboration}
% \SC{"Thus, Sigstore is a major identity-based software signing platform" currently sounds like it is identity-based because it is popular--I don't think that's what we're trying to communicate with the idea of next gen?}\KC{any better?}

\subsubsection{Sigstore Components \& Workflow}
Following Newman \etal~\cite{newman_sigstore_2022}, we outline \sigs architecture, components, and workflow to orient the reader.

\begin{itemize}[leftmargin=12pt, rightmargin=5pt]
  \item \textbf{Cosign:} Command-line and library support for creating, verifying, and bundling signatures and attestations. Integrations may differ across ecosystems (\eg GitHub Actions vs. Maven plugins). %, or other custom tooling).
    %\ST{I wonder if this should be generalized (e.g., github doesn't use cosign but uses the gh cli, I believe maven e.g., uses sigstore-java + some custom plugin arch}
  \item \textbf{Identity Provider:}
Services such as GitHub or Google that authenticate signers and issue short-lived identity tokens (\eg via OpenID Connect). %, OIDC, though other approaches like Decentralized Identifiers are also possible)\ST{OIDC identity provider is not only a RAS Syndrome, but IdP is likely more complete and OIDC is just an impl detail (we have a DiD/CryptID dispatch variant f.e.,)}
  \item \textbf{Fulcio (Certificate Authority):}
    Issues ephemeral signing certificates after verifying the signer’s OIDC token. %, eliminating the need for long-term key management.
  \item \textbf{Rekor (Transparency Log):}
    Append-only ledger that records each signing event and certificate issuance. %, enabling public auditing of signature provenance.
  \item \textbf{Monitors and Verifiers:}
    Tools that validate the signer's identity, certificate status, and signature integrity.
\end{itemize}

\begin{figure}
    \centering
    \includegraphics[width=0.90\linewidth]{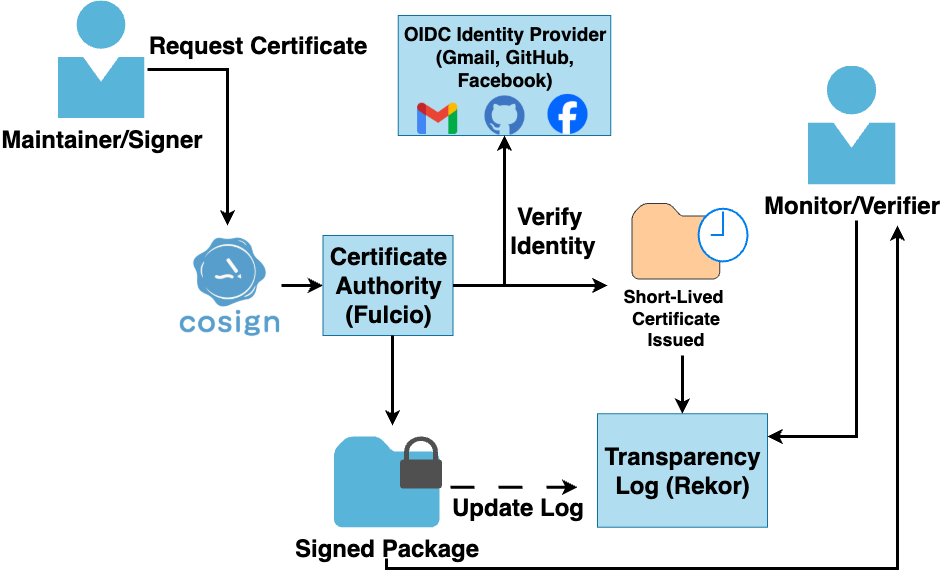}
    \vspace{5pt}
    \caption{
    Sigstore Signing Workflow. The software author requests a certificate from a certificate authority, which confirms the signer's identity through an identity provider. The signature and signer's identity are recorded in a transparency log, which the verifier can monitor to confirm the validity of the signature upon downloading the signed package.
%    \ifEXTENDED
%    See~\cref{sec: appendix-LegibleImage} for legible figure
%    \else
%    See~\cref{sec:DataAvailability} for legible figure
%    \fi
}
    \label{fig: sigstore_fig}
    %\vspace{-60pt}
\end{figure}

\cref{fig: sigstore_fig} summarizes the Sigstore software signing workflow.
First, the signer authenticates with an OIDC provider to obtain a short-lived token.
Next, the certificate authority Fulcio verifies this token and issues a one-time signing certificate.
The signer then uses Cosign to apply that certificate to the software artifact.
Finally, Rekor logs the certificate and signature, and verifiers retrieve entries from both Fulcio’s log and Rekor to confirm the signer’s identity and ensure the certificate and signature remain valid.  

%Although designed to operate as an integrated system, each component can be adopted independently—for example, using Rekor as a standalone transparency log for an existing CA, or integrating only Cosign and Fulcio into custom CI/CD pipelines. It can also be deployed within a private infrastructure rather than relying on the public instance.

\subsubsection{Commonality with other Identity-Based Tools}
% \JD{I'm looking for explicit connection to the phrase `generalizability of this case study'. Maybe just give a cref to Threats/Limitations later on?}
Sigstore, like other identity-based signing tools such as OpenPubKey~\cite{heilman_openpubkey_2023}, AWS Signer~\cite{aws2024signer}, and SignServer~\cite{signserver_about_2025}, shares core functionality in signer identity management and automated key handling. 
However, unlike Sigstore, most of these tools do not include a built‐in certificate authority (\eg OpenPubKey) or a transparency log (\eg OpenPubKey, SignServer); instead, such services must be added via external integrations.
We therefore use Sigstore as a focal case, while expecting many of our observations to transfer to similar identity-based signing tools.
% This shared design philosophy suggests that usability concerns may potentially generalize 
% While our focus on Sigstore allows for an in‐depth case analysis, many of its design principles—such as identity‐based authentication and ephemeral keys—are shared by its peers, suggesting potential generalizability across next‐generation signing tools.

% Sigstore exemplifies next‐generation, identity‐based signing with its seamless OIDC integration and bundled transparency log, many of its core features—short‐lived certificates, external identity providers, and append‐only audit logs—are shared across this emerging class of tools (e.g., OpenPubKey, SignServer). 
% Sigstore’s particular combination of these elements and its tight CI/CD integrations, however, distinguish it as a leading, production‐ready implementation in the broader ecosystem.  

\subsection{Instrument Design \& Development}
\label{sec:Method-InstrumentDesign}

%Our research objectives explore the impact of \textit{Software Tooling} as a factor in the practice of software signing from the viewpoint of practitioners, with a particular focus on how the usability of \signs tools influences their adoption.
\iffalse
Our study as explained in \cref{sec: bg_rw}, is a formative usability study and as described by Lewis\cite{lewis2014usability}, this type of evaluation offers aflexibility in its evaluation methods prioritizing expert practioitner inputs to better articularte the usability concerns of the system.
Following grecommendations from Salda\~na~\cite{saldana2011fundamentals}, we used semi-structured interviews to examine various aspects of tooling usability, \eg implementation challenges, perceived benefits, and comparison to other signing tools.
This approach let us pose consistent questions across all subjects, with the ability to probe about unique aspects of each subject's circumstances.
\fi
Per~\cref{sec:Background-Usability},
our study is a \emph{formative usability} study.
This type of evaluation offers methodological flexibility and prioritizes input from experienced practitioners to better articulate a system’s usability concerns and produce actionable guidance~\cite{lewis2014usability}. 
To elicit long-form, context-rich opinions and detailed observations, we employed semi-structured interviews to examine various aspects of tooling usability (\eg implementation challenges, perceived benefits, and comparisons to other signing tools), following guidelines from Salda\~na~\cite{saldana2011fundamentals}. This approach allowed us to pose a consistent core of questions across all participants while probing unique aspects of each participant's circumstances.

Sigstore is an emerging (though maturing) technology, and case study research recommends exploratory designs in such contexts~\cite{yin_case_2018, Runeson_Höst_2009}.
To develop our interview protocol, we therefore pursued open, inductive inquiry to capture the full range of practitioner usage experiences.
After this exploratory phase, we mapped our findings onto a formal usability framework, ensuring it reflected practitioners' experiences rather than constraining data collection to preconceived categories~(\cref{sec:Method-DataAnalysis}). % (see \cref{sec:Method-UsabilityFramework}). 
Therefore, in constructing the protocol, we: 

\begin{enumerate}[leftmargin=*]
  \item Reviewed grey literature, Sigstore blogs~\cite{sigstore_dev} and usage reports~\cite{Blauzvern_2022, Vaughan-Nichols_2022_b, Hutchings_2021, Sigstore_2022a, Sigstore_2022b, Sigstore_2022c, The_Sigstore_Technical_Steering_Committee_2023}, to ground our questions in real‐world implementations and updates.
  \item Used a snowball review of top papers on signing primitives~\cite{newman_sigstore_2022,merrill2023speranza} and relevant empirical studies (see~\cref{sec:Background}) to identify gaps and avoid redundant questions. We seeded the review with works from prominent cybersecurity and software engineering venues (\eg USENIX Security, IEEE S\&P, ICSE, FSE) published in the past 10 years.

  % \item Aligned questions with our formative usability framework (see~\cref{sec: knowledge_gaps}), with categories D3–D4 probing current tooling experiences and D1–D2, D5 exploring previous tools and transition factors (see~\cref{tab:InterviewProtocolSummary} for summary of our entire interview protocol).
  % \item Examined grey literature—Sigstore blogs~\cite{sigstore_dev} and usage reports~\cite{Blauzvern_2022, Vaughan-Nichols_2022, Hutchings_2021, Kammel_2022}—to ground our questions in real‐world implementations and updates.
  % \item Conducted a snowball review of security‐conference works on signing primitives~\cite{newman_sigstore_2022, merrill2023speranza} and empirical signing studies (see~\cref{sec:bg_rw}) to identify knowledge gaps and avoid redundancy.
\end{enumerate}

Next, we refined our interview protocols through a series of initial interviews~\cite{chenail2011interviewing}.
We first conducted practice interviews with the secondary authors of this work, followed by pilot interviews with the first two interview subjects.  We modified our protocol after practice and pilot, affecting 8 questions.\ifARXIV\footnote{Details are in ~\cref{sec:appendix-InterviewProtocol}.}\else\footnote{Protocol revisions are summarized in the technical report appendix; see ~\cref{sec:appendix-TechnicalReport}.}\fi.

\iffalse
For example, we moved the question about each participant's team's major product to the demographics section to provide context for their responses.
\fi
% Our fullinterview instruments  consisted of four main topics (A-D). Topics B and C were developed using a slightly different process than the ones described here and they are analyzed in another paper [BLINDED]. This exclusion followed best practices outlined by Voils \etal ~\cite{voils2007methodological}, as both parts did not fit within a similar analysis framework. Topic Topic D consisted of Five main question categories that are conducted in a semi structure manner.
% were developed using a slightly different process than the one described here and
\iffalse
Excluding the demographic section, Our full interview instrument consisted of two main topics (B \& C). 
Topics B consists of contextual questions on what risks prompts signing and how is signgn implemented across the software production process. These were different from demographics as they were not classification questions but ...
Topic C presents the questions related to assessing the usability of the signing tool option used by the participant's organization.
\fi
%are analyzed in another work [BLINDED].
%This exclusion follows best practices outlined by Voils et al. ~\cite{voils2007methodological}, as both parts did not fit within a similar analysis framework.

\cref{tab:InterviewProtocolSummary} summarizes our final interview protocol\footnote{Questions in the protocol reflect our screening criteria (expert users of signing infrastructure), with later questions building on earlier answers.}. 
After the demographic section (Topic A), our interview instrument consisted of two main topics (B \& C). 
\textbf{Topic B} comprised contextual questions about the supply chain risks that prompt signing and how signing is implemented across the software production process (\eg where signing is required vs.\ optional, which artifacts are signed, and relevant workflow/policy constraints).
%Unlike demographics, these were not classification questions; they were open-ended prompts intended to elicit organizational practices and context. 
\textbf{Topic C} presented questions to assess the usability of the signing tool(s) used by the participant’s organization, including perceived strengths and weaknesses, adoption rationale, implementation details, challenges, and any considerations about switching tools.

{
\renewcommand{\arraystretch}{1.3}
\begin{table}
    \centering
    \caption{
     Summary of interview protocol.
    %The decision to analyze these parts of the protocol separately was made in accordance with best practices.
    See \ifARXIV \cref{sec:appendix-InterviewProtocol} \else \cref{sec:OpenScience} \fi for the full protocol. 
    % \JD{ASE doesn't have appendices, so this is in the Artifact right?}
%    \JD{Having done that, we also need to introduce the contextual element along with the Demographics. Don't plagiarize, but repeat the redundant aspects from USENIX. Or possibly you'll see new aspects to report here that didn't make sense in USENIX.}
   }
   \vspace{5pt}
   \small
   \scriptsize
    \begin{tabular}{
    p{0.27\linewidth}|p{0.63\linewidth}
    }
    \toprule
        \textbf{Topic} & \textbf{Sample Questions} \\
    \toprule
         A. Demographics & What is your role in your team?  \\
         \midrule
         B. Contextual Information (Risks \& Use of Signing) & Can you describe any specific software supply chain attacks (\eg incidents with 3rd-party dependencies, code contributors, OSS) your team has encountered? How does the team use software signing to protect its source code (what parts of the process is signing required)? \\
         \midrule
         C. Signing Tool Usability Questions  &  \\
    % \midrule
    %     \multicolumn{2}{l}{\textbf{Question Categories in Topic C}} \\
    % \midrule
        C1. & What factors did the team consider before adopting this tool/method (Sigstore) over others? \\
        C2. & What was the team’s previous signing practice/tool before the introduction of Sigstore? \\
        C3. & How does your team implement Sigstore (which components of Sigstore does the team mostly use)? \\
        C4. & Have you encountered any challenge(s) using this tool of choice? \\
        C5. & Have you/your team considered switching Sigstore for another tool? \\
    \bottomrule
    \end{tabular}
    \label{tab:InterviewProtocolSummary}
\end{table}

}

\subsection{Data Collection}
\label{sec:Method-DataCollection}

\myparagraph{Population} \label{sec:population}
% \JD{Refactor `recruitment' so that we talk about population, then recruitment.}
% \myparagraph{Recruitment} \label{sec:recruitment}
% Our study entails the investigation of signing tools with respect to the \sig ecosystem. 
% As such, our target participant pool is centered on \sig users. Additionally, Our research questions as reflected in our interview instrument, reflects questions that aims to understand how software teams decide on software tools (\sig), how \sig is implemented, problems and advantages of \sig. We deduce that to effectively answer questions on decisions-making we required a target population of high-ranking or experienced practitioners, who either owned the compliance of \sign, or are in charge of the infrastructure of their respective teams/organizations.
% To effectively recruit this target population, we employed a non-probability-
% based purposive and snowball sampling approach to recruit participants ~\cite{baltes_sampling_2022}.
% We relied on the personal network of one of the authors to send out an initial invitation to members of the 2023 Kubecon (hosted by Linux foundation) Organizing committee, who were also part of the Intoto~\cite{in-toto_The-Linux-Foundation} Steering committee. 
% This initial invitation yielded XX participants. We then snowballed from the recommendation of other participants from the initaila pool of participants to resruit other XX participants.
Our study investigates signing tools using a case study of the \sig tool ecosystem.
% As such, our target participant pool is centered on \sig and potential \sig users.
Accordingly, our participant pool focused on current and prospective users of this tool.
% \SC{\sig and potential \sig users?}
% \JD{looks like a good change}\KC{good?}
% Furthermore, our research questions, as reflected in our interview instrument, aim to understand the reasons prior to adoption that teams choose software tools (\sig),how \sig is used in signing, and the challenges and benefits of \sig.
Furthermore, since our research questions pertain to organizational decision-making, %satisfactory answers to our research questions %, as reflected in the interview instrument, aim to understand the factors influencing adoption decisions, how the tool is used in practice, and the challenges and benefits experienced by teams.
%To effectively answer these questions related to decision-making,
we targeted expert practitioners who either oversee \signs compliance or manage the infrastructure of their respective teams or organizations.

% \JD{Recall an issue about positionality --- Do we have a Statement of Positionality in this paper? If not please consider.}
To recruit this target population, we employed a non-probability purposive and snowball sampling approach ~\cite{baltes_sampling_2022}.
We leveraged the personal network of one of the authors to send an initial invitation to members of the 2023 KubeCon (hosted by the Linux Foundation) organizing committee, who were also part of the In-toto~\cite{intoto, in-toto_Linux_Foundation} Steering Committee (ITSC).
This initial invitation yielded 6 participants.
We then expanded our recruitment through snowball sampling, relying on recommendations from initial participants (4) and authors' contacts (7) to recruit an additional 11 participants.
We report on data from 17 participants.\footnote{We conducted 18 interviews. Participant 18 was a senior security expert. While aware of the organization’s overall use of signing, they reported limited knowledge of the signing tool's specific operation and deployment.} % Accordingly, we did use this data.} %not rely on this interview for deployment-specific analysis and report results for 17 participants.}

 % framework.\footnote{The other subject subject provided limited information about their \signs tool.}
% Two of them used \signs tools other than \sig, and three gave limited information about their \signs tool.

\iffalse
To recruit our survey participants, we employed a targeted sampling approach ~\cite{watters1989targeted}. We first focused on \sig users by targeting participants of the 2024 SigstoreCon conference (co-hosted with the 2024 KubeCon conference) ~\cite{sigstorecon2024}. Additionally, through the professional network of one of the authors, we reached out to the broader \sig user community via post-conference communications. This sampling approach yielded YY survey participants.
\fi

\myparagraph{Participant Demographics}
Our subjects were experienced security practitioners responsible for initiating or implementing their organization's security controls or compliance.
They thus had relevant context to assess their organization’s strategies for adopting \signs tools. 
Subjects came from 13 distinct organizations, all companies.
Relevant demographics are in~\cref{tab:Subjects}.
Thirteen participants reported their organizations use Sigstore.
Two use internally developed tools and two rely on Notary v1 and PGP signing.
Those four participants enrich our findings, as their experiences shed light on usability factors that \textit{hinder} adoption of Sigstore by non-users.

{
	%\arraystretch{1.5}
	\begin{table}[!t]
		\centering
        \caption{
        Demographics.
        %For anonymity, we used generic job roles, not specific titles.
        For \ul{Role}, ``\textit{Senior management}'' are senior managers/directors/executives;
        ``\textit{Technical leader}'' are senior/lead/partner/principal engineers;
        and
        ``\textit{Engineer}'' and ``\textit{Manager}'' are junior staff. \ifARXIV \else We add interview duration to an extended version of this table (\cref{sec:appendix-TechnicalReport}).\fi 
        \cref{tab:orgSummary} gives organizational info.
        }
        \vspace{5pt}
        \scriptsize
        % \small
		\begin{tabular}{llcl}
            \toprule
           \textbf{ID} & \textbf{Role} & \textbf{Experience}  & \textbf{Software Type} \\
            \toprule
            P1 & Research leader & 5 years  & Internal POC software \tablefootnote{POC: Proof of concept software: early-stage software that may be productized.} \\
            P2 & Senior mgmt. & 15 years  & SAAS security tool \\
            P3 & Senior mgmt. & 13 years  & SAAS security tool \\
            P4 & Technical leader  & 20 years & Open-source tooling \\
            P5 & Engineer   & 2 years & Internal security tooling  \\
            P6  & Technical leader & 27 years & Internal security tooling \\
            P7 & Manager  & 6 years  & Security tooling \\
            P8 & Technical leader  & 8 years & Internal security tooling \\
            P9 & Engineer & 2.5 years  & SAAS security \\
            P10 & Engineer & 13 years & SAAS security \\
            P11 & Technical leader & 16 years & Firmware \\
            P12 & Technical leader & 4 years  & Consultancy \\
            P13 & Senior mgmt.  & 16 years & Internal security tool \\
            P14 & Research leader & 13 years  & POC security software \\
            P15 & Senior mgmt.  & 15 years & Internal security tooling \\
            P16 & Senior mgmt.  & 15 years & SAAS \\
            P17 & Manager  & 11 years  & Security tooling \\
            \bottomrule
		\end{tabular}
		\label{tab:Subjects}
	\end{table}
}

% This involvement of our subjects in the operational and security decision-making gave them the relevant context to assess their organization’s adoption of \signs tools.
% Not all of our participants directly oversaw compliance, however, their involvement in security strategy and operational decision-making provided them with the appropriate context to assess the decision-making details of their organization’s adoption of \signs tools.
\myparagraph{Interviews}
We conducted our interviews over Zoom.
% \JD{See the fill-in in the next sentence --- FRACTION}
Each interview lasted $\sim$50 minutes. 
% with the parts pertinent to topic D comprising $\sim$15 minutes (one third) of each interview.\JD{bill the whole interview time, not just 15 mins}
The lead author conducted these interviews.
% We offered each subject a \$100 gift card as an incentive, in recognition of their experience level. 
We offered each subject a \$100 gift card as an incentive in recognition of their expertise.

% \myparagraph{Survey Attempt}
% We attempted a survey (details are in ~\cref{sec:appendix-SurveyAttempt}) but received only a few completed responses, echoing the challenges of surveying specialized communities~\cite{steeh1981trends}.

\myparagraph{Survey Attempt}
We recognize the importance of data triangulation in case studies and therefore fielded a follow-on survey to complement our interviews.
%Approximately one year after the interview phase,
We deployed a survey to validate and model the prevalence and stability of our interview themes over time and across contexts. We tested our survey instrument in two practice runs with two members of our research team.
Despite five months of repeated solicitation in the \sig community, we received only 13 completed responses, of which 7 were sufficiently complete for analysis (\ifARXIV see~\cref{sec:OpenScience}\else see~\cref{sec:OpenScience}\fi).
This regrettably echoes the well-documented difficulty of recruiting participants from highly specialized populations~\cite{steeh1981trends}.
%Given the limited response, we report our survey results descriptively and do not integrate it into our core analysis.

Data triangulation is encouraged but not mandatory in case study research. %, it is not mandatory.
%method choices are also constrained by time, resources, and access.
As Stake notes, combining multiple data sources strengthens the substantiation of constructs and hypotheses, yet practical constraints necessarily shape feasible designs~\cite{Stake1995ArtCaseStudyResearch}. 
We encountered such constraints. %, most notably access to sufficient system users.
% Our study faced such constraints—most notably access—consistent with prior software engineering case studies that have employed single-source interview or survey designs~\cite{davila2024,pagano2013}.

\iffalse
Our surveys were distributed via Qualtrics and were designed to take approximately 7 minutes to complete.
\fi

\subsection{Data Analysis}
\label{sec:Method-DataAnalysis}

\ifEXTENDED
We transcribed our interview recordings using \url{www.rev.com}'s human transcription service and anonymized all potentially identifiable information.
\fi

We analyzed data in two stages:
  an interpretive–exploratory thematic analysis,
  and then
  a retrospective mapping of emergent themes onto a formative usability framework.

% To analyze our data, \ul{first}, we used an \emph{interpretive}~\cite{Easterbrook_Singer_Storey_Damian_2008}
% and \emph{exploratory} approach for initial analysis, as recommended for studies characterized by substantial uncertainty~\cite{rowlands2005grounded}, in our case, uncertainty around how organizations adopt software signing as a security method, a decision that varies with organizational policies, security expectations, perceived threats, and customer requirements.
First, we adopted an \emph{interpretive} and 
\emph{exploratory}~\cite{Easterbrook_Singer_Storey_Damian_2008} approach for initial analysis, as recommended for 
studies characterized by substantial uncertainty~\cite{rowlands2005grounded}. In our case, this uncertainty concerned how 
organizations adopt software signing as a security method, a decision that varies with 
organizational policies, security expectations, perceived threats, and customer requirements. 
% To analyze our data, \ul{first}, we used an \emph{interpretive}~\cite {Easterbrook_Singer_Storey_Damian_2008}
% and \textit{exploratory} approach for initial data analysis, as recommended for studies characterized by uncertainty~\cite{rowlands2005grounded}.
This orientation motivated the use of a reflexive thematic analysis lens~\cite{braun_using_2006}. At the same time, we aimed to generate concrete, usable guidance for tool designers. To increase the systematicity and transparency of our process, we therefore incorporated selected coding-reliability practices, structured early coding, and agreement checks~\cite{oconnor_intercoder_2020}. 
In short, reflexivity grounds our insights, and these reliability practices help make them more communicable and actionable.
%Here, identity-based software signing tools constitute a new class that differs substantially from legacy key-managed tools and has not been examined in this context.

Second, because our work is a usability study, we situated the exploratory findings in a usability context by applying a formative usability framework retrospectively to our inductively generated results.
This retrospective mapping of emergent themes onto existing constructs is an established interpretive strategy in qualitative analysis.
%variants include framework-informed thematic analysis, the Framework Method, template analysis, and best-fit framework synthesis. These approaches 
This method uses prior constructs to structure and extend inductively derived findings while preserving empirical grounding (\eg \cite{Fereday2006, Gale2013FrameworkMethod, Carroll2013BestFitFramework}).

\iffalse
To outline our process, \textbf{Stage 1—Exploratory (Interpretive)} comprised (1) familiarization with transcripts~\cite{terry2017thematic}; (2) inductive memoing and coding, developing initial codebooks without reference to any usability framework~\cite{Fereday2006}; and (3) theme development from the inductively generated codes. \textbf{Stage 2—Formative Usability Framing} consisted of (4) a retrospective mapping of themes to a formative usability conceptual model to refine interpretation in a usability context (see \cref{fig:methodology}). We describe these steps in detail next.
\fi

\paragraph{Memoing \& Codebook Creation.}
Two analysts participated in this analysis stage.
After transcription\footnote{Interview recordings were transcribed by www.rev.com using human transcription to address spoken accents.} and anonymization, we began with data familiarization~\cite{terry2017thematic}.
During Familiarization, the main analyst\footnote{The lead author is the main analyst in this study.} read all interviews, and the secondary analyst read three transcripts to calibrate.
% We began with the initial 17 transcripts.

Next, the two analysts independently memoed a subset of 6 anonymized transcripts through three rounds of review (2 transcripts per round) and discussion, following the recommendations of O'Connor \& Joffe~\cite{oconnor_intercoder_2020} and Campbell \etal~\cite{campbell2013coding}. This process improved reliability of the analysis while making efficient use of resources.\footnote{Although there is no consensus on the ideal proportion of data to use, O’Connor and Joffe note that selecting 10--25\% is typical~\cite{oconnor_intercoder_2020}. We randomly selected six transcripts ( $\sim$35\% of our dataset) to develop our initial codebook.} %, using this slightly higher proportion to support cross-analyst calibration of memoing conventions.}
% \SC{make this active voice instead}\KC{done}
After each round, we discussed emerging codes in meetings, which produced 506 coded memos in total.
We distilled these memos into a codebook (36 categories) over several multi-hour meetings.

Next, to assess the reliability of the coding process, we combined the techniques outlined by Maxam \& Davis~\cite{maxam2024interview} and Campbell \etal~\cite{campbell2013coding}. Specifically, we randomly selected an unanalyzed transcript, which we coded independently. 
Following Feng \etals~\cite{feng_intercoder_2014} recommendation, we used the percentage agreement measure\footnote{Calculated as: \\ \hspace*{1cm}\textit{(matching code assignments / total code assignments) × 100}} to evaluate consistency, as both analysts had contributed to developing the codebook.
We obtained an 89\% agreement score, suggesting a stable coding scheme~\cite{campbell2013coding, oconnor_intercoder_2020}.
We resolved disagreements by discussion. 
% This high agreement according to Campbell \etal implies that our coding scheme was stable enough to use in singly coding remaining uncoded transcripts.
% These were subsequently reviewed by analysts A1 and A3.

\paragraph{Codebook Revision.}
% Following the reliability assessment of our codebook,
% Since the agreement recorded was substantive, Analyst A1 proceeded to code the remaining transcripts in a set of 2 batches of 6 transcripts per each round.  For the first selected 6 transcripts, this phase added 6 new code categories. This was reviewed by Analyst A3.
% The same process of analysis was applied to the final 6 transcripts. This added 4 more code categories tot he codebook as was also reviewed by analysts A3. We note that of the 10 codes added in this phase of rveising the codebook, only one 3 provided new insights into our data, thus By agreements of A1 and A3, we categorized these as minor categories.
Following the reliability assessment of our codebook and given the substantive agreement recorded, the main analyst proceeded to code the remaining transcripts in two batches of six and five transcripts per round, respectively. The lead analysts revisited earlier coded transcripts throughout. In the first batch, six new code categories were introduced, which were subsequently reviewed by the other analyst.
% The same analytical process was applied to the final six transcripts, resulting in the addition of four more code categories, which were also reviewed by the other analyst. 
The same analytical process was applied to the final five transcripts, leading to the addition of four new code categories, which were subsequently reviewed by the other analyst.
Notably, of the 10 codes added during this phase of codebook refinement, only three provided new insights into the data.
By mutual agreement between both analysts, these were categorized as minor codes.

To assess the adequacy of our sample size and analysis, we calculated code saturation for each category in our codebook.
Following the recommendations of Guest \etal~\cite{guest2006many}, we measured the cumulative number of new codes introduced in each interview (\cref{fig: saturation}).
We saw saturation after interview \#12. %, with no new codes after that.
Each interview yielded a median of 15 codes.
% \ifEXTENDED
% The results are presented in~\cref{fig: saturation}.
% \else
% See~\cref{sec:DataAvailability} for saturation curve.
% \fi

% aturation was measured after we concluded the analysis of all 18 interview transcripts. We followed Guest \etals ~\cite{guest2006many} recommendation to cumulatively measure the number of new codes appearing in each interview. Following Maxam \etals ~\cite{maxam2024interview}, we present a plot of this in~\cref{fig: saturation}. We can see that saturation was achieved after interview 11, as no new code was observed. Each interview had a median number of unique codes, 33 as shown in the same~\cref{fig: saturation}.

\paragraph{Thematic Analysis \& Usability Analysis}
\label{sec: thematic_usability_analysis}

\iffalse
\myparagraph{Thematic Analysis}
% Each interview had a median of 15 codes.
% Next, following the method proposed by Braune \& Clark \cite{braun_using_2006} we induced initial themes from our code categories. We focused on inducing themes related to the D part of our protocol as we highlighted earlier~\cref{tab:InterviewProtocolSummary}. 
% We performed Another round of thematic analysis using the themes from the initial thematic analysis and our formative usability framework \cref{sec: knowledge_gaps}. We embedded our initial themes within the factors defined in our frameworks.
Next, following the method proposed by Braun \& Clarke \cite{braun_using_2006}, we derived initial themes from our code categories.
% We focused on identifying themes related to Topic D of our protocol 
(\cref{tab:InterviewProtocolSummary}).
\fi

% \myparagraph{Thematic Analysis}
 Next, following Braun \& Clarke~\cite{braun_using_2006}, we derived initial themes from our code categories (see \cref{tab:InterviewProtocolSummary} for protocol focus). This reflexive thematic analysis provided an inductive account of participants’ experiences and practices. 
The lead analyst generated the themes and then discussed and refined them with feedback from the secondary analyst.

From the emergent themes, we interpreted our findings using a formative usability framework. 
Lacking formative usability studies set in similar contexts in the Software Engineering literature, we followed advice~\cite{lorey_social_2022, Fernandez2019EmpiricalSEInterdiscipline, Sjoberg2007FutureEmpiricalMethods}
to seek theory from adjacent disciplines.
Surveying candidate frameworks beyond SE,
we found three groups:
 formative usability assessments of software systems~\cite{chan2022practical};
 applications of technology systems in other disciplines~\cite{strifler2024development,shah2023my, cresswell2020developing};
 and
 general usability frameworks~\cite{andre2000testing}.
Among these, we selected Cresswell \etals four-factor TPOM framework (\textit{Technology, People, Organization, Macro-environment})~\cite{cresswell2020developing} since it best aligned with our inductively generated themes (see \cref{fig:methodology}). Our intent was to synthesize those themes into higher-level conceptual groupings that correspond to established usability determinants of adoption. The TPOM categories also suggest interventions, \eg People issues may not be resolved via Technology. %suggest interventions in community, ecosystem, or regulatory engagement rather than purely technical changes.

We then conducted a framework-informed analysis, mapping the emergent themes to the TPOM factors to refine and structure our interpretations. % within a formative usability perspective.
The lead analyst mapped our earlier generated themes and then discussed and refined the mappings with feedback from the secondary analyst.

\section{Results}
\label{sec:Results}

\begin{figure}
    \centering
    \includegraphics[width=0.98\linewidth]{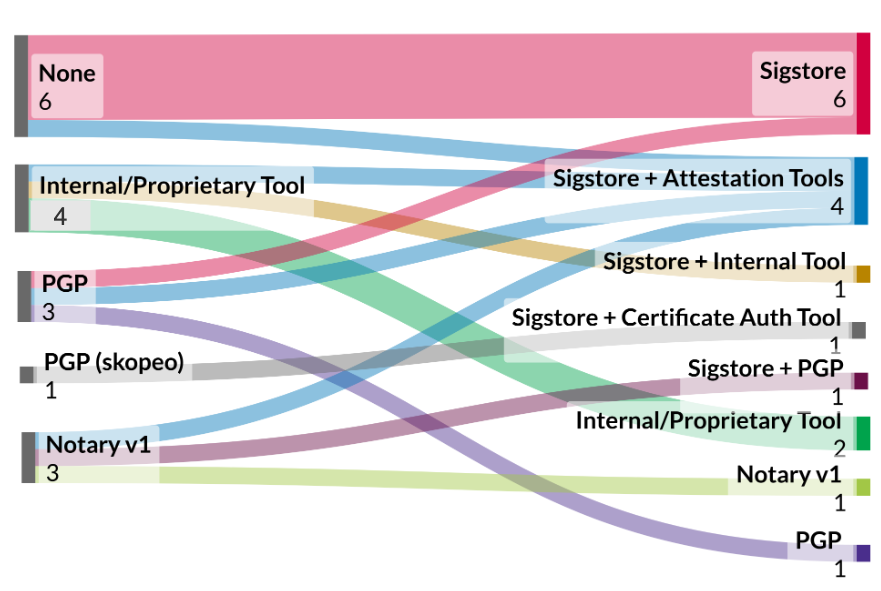}
    \caption{
   Self-Reported Evolution of Software Signing Tools.
    Sankey diagram showing participants’ self-reported transitions between previously used and current signing tools in their organizations. Flows (e.g., from no signing to Sigstore) reflect participant accounts, not our observations.
    %Most reported their organizations moved from no former tools, a pngform of GPG signing or notary to \sig.
    %Most participants depicted with the flow from none to a sigstore implementation report that; they either initially used no tools, or have always used Sigstore.
    \JD{We are missing one label on the right side (above green path)}
    }
    \label{fig:sankey-flow}
\end{figure}

% First, we present a sankey flow diagram of our participants teams use of \signs tools up to present. While a sizeable portion of our participants do not mention a shift (6) in \signs tools used( most do not mention any, some mentioned that they still use their initial \signs tools and only currently combine them with \sig). The next major shift was from GPG based signing implementations (3) to Sigstore, and Notary (2) was another shift reported shift. Other shifts mentioned was Skopeo (1) unspecified internal tools(1) and a proprietary tool(1).
% \SC{feels like abrupt start to section}\KC{better?}\SC{yep!}
\iffalse
First, we present a Sankey diagram illustrating the changes in signing tool usage among our participants' teams (\cref{fig:sankey-flow}). 
\fi
We begin by presenting a Sankey diagram (\cref{fig:sankey-flow}) that illustrates the distribution of our participants between Sigstore and non-Sigstore users, and how their teams changed signing tool usage over time.
% A significant portion of participants (6) depicted as moving from none to adopting \sig did not report a shift in signing tools—either they did not specify any changes (either moved from using no tools or have been using Sigstore since inception) or stated that they continue using their initial signing tools while integrating them with \sig. 
Many participants (6) transitioned from using no tools to \sig. % --- either their organization has always used \sig organization or continued using their initial signing tools alongside \sig.
Others transitioned from existing tools to \sig:
3 from implementations that had some combinations of GPG/PGP (of which one practitioner reported combining PGP with Skopeo\cite{RedHat2022WhatIsSkopeo}), 2 from Notary, and 2 from proprietary or internal tools. 
%The most notable transition was from GPG-based signing implementations (3) to \sig, followed by a shift from Notary (2). Other reported transitions included Skopeo (1), unspecified internal tools (1), and a proprietary tool (1).
4 subjects did \textit{not} adopt Sigstore.

%We distinguish our results by RQ
To understand the usability concerns that Sigstore adopters face; and why practitioners did (and did not) change tools,
we structure the results by RQ using Cresswell's framework.

% We interviewed 18 expert practitioners, 12 of these reported some experience in working with sigstore. These experiences range from practitioners who have set up sigstore as a part of their build pipeline, in test capacity, and as a part of other signing implementation.
% We categorize our preliminary findings into the following themes.
% a.	Problems experienced with tooling
% b.	Advantages of tooling
% c.	Reason for adopting tooling
% d.	Change in adopted tooling
% e.	Manner of tool implementation
% Additionally in our methodology, we conduct a survey to validate the findings we have elicited from our subjects. Our survey questions are mapped to our 3 research questions, but they are also formulated to to investigate the 6 major themes our interview thematic analysis yields.
\subsection{RQ1: Strengths \& Weaknesses of Sigstore}
\label{sec:Results-RQ1}

The first aim of our research work was to ascertain from the experiences of our subjects the strengths and weaknesses they have experienced while using \sig. 
%We aligned each identified theme to factors from our adopted usability framework: Technological Factors – situations where observed conditions were linked to the behavior of Sigstore technology itself, Human/Social Factors – participants' observations related to gaps in support resources, previous experiences (with other tools) Organizational Factors – aspects tied to users' organizations or their organizational capacities, and Macroenvironmental Factors – external influences affecting adoption and use.

\subsubsection{\sig Strengths}
\label{sec: sigstore_strength}
%We present a summarized list of practitioner-reported advantages of using Sigstore in~\cref{tab:rq1a}.
% First, we extracted from our participants' responses the benefits they have experienced while using sigstore. These responses reflects our participants experiences while using \sig.  
% Their responses is grouped under 8 topics -- Ease of use, Identity management for Signers, \sigs compatibility with several new technologies, \sigs use of Short-lived keys and certificates, \sigs implementation of a transparency log, \sigs encouragement of bundling signatures with attestations, Free cost for using the public \sig instance, and Reliability of the \sig service.
We extracted the benefits our participants experienced while using Sigstore from their responses.
These reflections capture their firsthand experiences with \sig.
% We grouped their responses into eight categories: ease of use, identity management for signers, compatibility with emerging technologies, use of short-lived keys and certificates, implementation of a transparency log, encouragement of bundling signatures with attestations, free access to the public \sig instance, and reliability of the \sig service.
We grouped their responses into eight categories as summarized in~\cref{tab:rq1a}.
When mapped to the Cresswell framework, these factors fell into two categories: \textit{Technological} and \textit{Macroenvironmental}.

{
	%\arraystretch{1.5} 
	\begin{table}
		\centering
		\caption{
		Summary of Practitioner-Reported Advantages of Using Sigstore.
        Practitioners cited \sigs \textit{technological} and \textit{macroenvironmental} factors as advantages.
            % \JD{Add another sentence giving interpretation.}
		}
        \vspace{5pt}
		\small
            \scriptsize
		\begin{tabular}{ p{0.65\linewidth}p{0.25\linewidth}}
            \toprule
            \textbf{Topics \& Associated Examples  } & \textbf{Subjects} \\
            \midrule
            \textbf{Technological Factors} & \\
            \textit{Ease of Use} & {\textit{8 subjects}} \\
             1. Signing Workflow \& Verification & P1, P2, P7, P9, P14-17 \\
             2. Setting up with automated CI/CD actions & P9 \\
             3. No key distribution problems & P1\\
            % \midrule
            \textit{Use of Short-lived Keys \& Certificate} & {\textit{3 subjects}} P2, P3, P15 \\
             % Use OIDC/Keyless signing &  \\
            % \midrule
            \textit{Signer ID Management} & {\textit{4 subjects}} \\
             1. Use of OIDC(Keyless) to authenticate signers & P3, P5, P10, P12 \\
             % Setting up with automated CI/CD actions & -- \\
             % No key distribution problems & --\\
            % \midrule
            \textit{Compatibility with Several New Technologies} & {\textit{4 subjects}} \\
             1. Integrability with SLSA build & P16\\
             2. Integrability with several container registries/technologies &  P12, P14 \\
             3. Integrability with several cloud-native applications & P15\\
            % \midrule
            \textit{Presence of a Transparency Log} & {\textit{3 subjects}} \\
             1. Transparency logs increase security & P5, P14 \\
             2. Evaluation of signing adoption using logs & P9 \\s            % \midrule
            \textit{Bundling Signatures With Provenance Attestations} & {\textit{2 subjects}} P3, P4 \\
             % Evaluation of organizational signing adoption & --- \\
            % \midrule
            \textit{Reliability of Service} & {\textit{1 subject}} P7 \\
            \midrule
            \textbf{Macroenvironmental Factors} & \\
            \textit{Free/Open-Source} & {\textit{2 subjects}} P7, P17 \\
              % & --- \\
             % Evaluation of organizational signing adoption & --- \\
            \bottomrule
		\end{tabular}
		\label{tab:rq1a}
	\end{table}
    
}

\paragraph{Technological Factors}
We identified five factors:

\myparagraph{Ease of use}
This was the most frequently discussed advantage of \sig over other tools.
Participants commonly highlighted the simplicity of the \sig workflow. %, particularly the automation of key management.
As \emph{\Subjectseven} put it, \myinlinequote{I don't really know of any other options that provide the same conveniences that Sigstore provides... It really is just one command to sign something and then one more command to verify it, and so much of the work is handled by Sigstore.} 
Other ease-of-use advantages mentioned were the seamless integration of the Sigstore workflow into CI/CD automation using GitHub Actions and the reduced complexity of key management.
% I can't really think of another public reliable service that's also free that makes it so easy not only to sign but then also for end users to verify. And it might be that I'm just not aware of what else is out there, but I really haven't seen anything else that is so easy to use from a user perspective.
% \ST{I'm left wanting a bit more on this bullet point. Do we have any data about *why* it is easy to use? is it key mgmt? safe defaults? I think here you may want to lean on previous literature of secure API design and other usability/adoption bits}

\myparagraph{Identity management for Signers}
% \sigs incorporation of the OIDc identity management capabilities was a key feature reported by our participants as a key benefits of their use.
% In legacy key-managed \signs the identity of the signer is mostly an ad-hoc feature that may or may not be required by the signing authority. In Sigstore this feature (implemented by the OIDC) is bundled with the Fulcio certificate authority as an identity management step prior to creating a signature. Participants greatly appreciated the truncation of the key generation (and distribution) phase of \signs that this affords them. Cutting away times to distribute to keys to key servers, searching keys on key servers etc. 
\sigs integration of OIDC identity management was a strength per our participants. In legacy key-managed \signs, signer identity is often an ad-hoc feature and may be optional for the signing authority. In Sigstore, however, this behavior is integrated with the Fulcio certificate authority as an identity management step before creating a signature. Participants greatly valued the elimination of the key generation and distribution phase, as it streamlined the signing process by reducing the time spent distributing keys to key servers and searching for keys. This aligns with the intuition that identity management is easier than key management for developers~\cite{newman_sigstore_2022}.
\emph{\Subjectsixteen} highlights this point:
  \myinlinequote{I can associate my identity with an OIDC identity as opposed to generat[ing] a key and keep[ing] track of that key...I could say, `Oh, this is signed with [Jane]'s public GitHub identity'...that's much better.} % So unless my GitHub identity has been compromised, that's much better.}
% \ST{Here, could you add something along the lines of ``this follows the intuition that identity management is easier than key management for developers~\cite{sigstore-paper}'' or so?}
% \JD{Yes, please}

% P10: they like a number of different things about Sigstore. They like that it uses OpenID Connect and it uses this centralized service that can be, in some sense, trusted because the organizations that are doing this, Google and Microsoft and others, they take security pretty seriously. So that itself actually can be a controversial notion about how much you should trust that, but I think these people tend to trust OpenID Connect and the OpenID Connect providers.

\myparagraph{Compatibility with Several New Technologies}
% A portion of our participants reports that \sigs integration and compatibility with a wide range of modern technologies offered them a great advantage. Among technologies mentioned by our participants ranged from other security toolings like SLSA build attestations, containerization technologies, cloud computing applications etc. The resource to setup connections to these technologies in most cases are minimal or none.
% \emph{\Subjectseven} says, \myinlinequote{}
A portion of our participants reported that \sigs integration and compatibility with a wide range of modern technologies provided a significant advantage. 
The mentioned technologies included security tooling such as SLSA build attestations, as well as containerization technologies and cloud computing applications. 
In most cases, the resources required to set up connections to these technologies were minimal or nonexistent.
\emph{\Subjecttwelve} states, \myinlinequote{a lot of clients are getting into signing their containers...
%pretty trivial to do if you use something like cosign...
Sigstore is the easy choice because it just works on any registry, and integrating it to Kubernetes is easy.}
% We discuss further in 
%\ST{I think it's very important here to drill a bit deeper, are there any indications that these findings can apply to tool design? how?}

% \ParticipantQuote{\Subjectthree}{... the other thing is policies, creating a policy that is, for one, it's hard to do. And then the other part is creating an effective policy and knowing that it's effective.}
\myparagraph{Use of Short-lived Keys/Certificates}
% \sigs enforcements of a short-period valid issued key/certificate  as a security feature was also acknowledged by pour participants as a major strength of their tooling.
% Prioir \signs tools requires users to set an expiration time to cryptographic variables like signatures, keys, and certificates when they are generated. while these gives the practitioners a customization ability, they present a possibility where cryptographic variables can remain valid for a long time, which increases the chance of their compromise. \sigs automated management of this, was appreciated by participants.
Participants appreciated \sigs enforcement of short-lived keys and certificates. % was recognized by our participants as a major strength of the tooling.
Some signing tools require users to manually set expiration times for cryptographic elements such as keys and certificates at the time of generation.
That customizability also introduces the risk of long-lived cryptographic materials, increasing the likelihood of compromise. 
Participants preferred \sigs automated management of key and certificate lifetimes.
As \emph{\Subjecttwo} stated, \myinlinequote{PGP, you still have to figure out the key distribution problem....Sigstore uses short-lived keys, that solves a lot of those issues.} % around key storage and key distribution...}

\myparagraph{Presence of a Transparency Log}
% The presence of a transparency log to serve as a \textit{tamper-resistant} secondary record of changes to an artifacts metadata in the \sig ecosystem was beneficial to participants.
% \sigs inclusion of the transparency log to the ecosystem provided users an extra layer of trust to verify signed artifacts. 
% The inclusion of a transparency log in the \sig ecosystem, serving as a \textit{tamper-resistant} secondary record of changes to an artifact's metadata, was seen as highly beneficial by participants. This feature provided an additional layer of trust, allowing users to verify signed artifacts with greater confidence.
Adopters valued the ability to audit signing actions.
\sigs transparency log provides a tamper-resistant record of all changes to an artifact’s metadata, a functionality not offered by other signing systems.
%This auditability gave users greater confidence in the integrity of signed artifacts.
\ParticipantQuote{\Subjectnine}{From a higher level, we're going to use a transparency log to determine what was actually signed and what was written into that transparency log.}
% And then you're able to verify those things within Sigstore ground record, that this is a transparency log to where you're going to be able to understand and verify what's been put in, what's been signed and who signed it.}

%Other technological factors noted by participants are the reliability of Sigstore’s public instance.
% \ST{I think here is where the ``marketting terms'' reaction arose from the SOUPS reviewers. Is there a way to rephrase this more along the lines of ``the ability to audit signing actions is crucial for adopters because...'' or ``no other signing systems we surveyed provide this ability to audit, which was appealing to adopters''}

\paragraph{Macroenvironmental Factors.}
% \vspace{3em}

\myparagraph{Free/Open-Source}
% \sigs open-source nature was also quoted as an advantage of using the tool by participants. Participants mentioned that the no-cost solution of \sig reduces the setup barrier, and maintenance resource barrier especially for users who use sigstore's public deployments.
% \ParticipantQuote{}{}
% In addition to the cost of setup and maintenance while using the public instance, the uptime of the \sig service was also quoted by \emph{\Subjectseven} to be reliable.
\sigs open-source nature was also highlighted by participants as a key advantage. They noted that this reduces both setup and maintenance barriers, particularly for users leveraging \sigs public deployments.
\ParticipantQuote{\Subjectseven}{I think it's amazing that Sigstore is free and public.}
This enthusiasm reflects both the appeal of free/open-source (FOSS) software and the practical benefits of public accessibility. 
However, note that open-source $\neq$ free; enterprises may pay for private deployments, support, and integration.

% Beyond cost considerations, participants also emphasized the reliability of Sigstore’s public instance, with \emph{\Subjectseven} specifically citing its consistent uptime as a valuable aspect of the service.
% \myparagraph{Reliability of Service}

\subsubsection{\sigs Weaknesses}
\label{sec: sigstore_weakness}
We summarize the difficulties reported by participants in using Sigstore in~\cref{tab:rq1_factor_view}.
As we detail, most of these weaknesses mapped to multiple usability factors.

{
\begin{table}[!t]
\centering
\caption{Summary of Practitioner-Reported Difficulties Using Sigstore.
Difficulties are grouped by Cresswell factors. %: Technology, Social/Human, Organizational (O), \& Macroenvironmental (M).
}
\vspace{5pt}
\small
\scriptsize
\begin{tabular}{p{0.65\linewidth}p{0.25\linewidth}}
\toprule
\textbf{Topics \& Associated Examples} & \textbf{Subjects} \\
\midrule
\textbf{Technological Factors} & \\
\textit{Transparency Log Constraints} & \textit{6 subjects} \\
1. Not suitable for private setup & P2, P3, P6, P14, P17 \\
2. Use in air-gapped/offline conditions & P2, P3, P9 \\
% \textbf{Air-gapped / Offline Operation} & \textbf{5} \\
% \quad Use in air-gapped conditions — \textit{\textbf{T}} & P2, P3, P9 \\
% \quad Offline capabilities — \textit{\textbf{T}} & P3, P4 \\
\textit{Performance \& Rate Limits} & \textit{6 subjects} \\
1. Rate limiting problems & P3, P7, P14, P15 \\
2. Latency concerns & P6 \\
\textit{Integrations \& Tooling} & \textit{3 subjects} \\
1. GitLab/Jenkins & P9 \\
2. Other unsupported technologies  & P16 \\
3. Attestation storage & P1 \\
\textit{Documentation (Technical Aspects)} & \textit{6 subjects} \\
% \quad Private instance setup documentation — \textit{\textbf{P \& T}} & P9, P12, P15 \\
1. Private instance setup documentation & P9, P12, P15 \\
% \textbf{Documentation \& Usage Clarity} & \textbf{3} \\
2. Other documentation / usage issues & P1, P14, P17 \\
\textit{Private Instance: Infra \& Cost} & \textit{2 subjects} \\
1. Infra requirements \& maintenance cost & P5, P6 \\
\textit{Fulcio / Timestamping Workflow} & \textit{1 subject} \\
1. Timestamping issues & P3 \\
2. Fulcio–OIDC workflow & P3 \\
\textit{Software Libraries} & {\textit{1 subject}} \\
1. Unsupported software libraries & P7 \\
\midrule
\textbf{Social / Human Factors} & \\[-2pt]
\textit{Log Monitoring Burden} & \textit{1 subject} \\
1. Effort to monitor logs & P2 \\
\textit{Support \& Maintenance} & \textit{3 subjects} \\
1. Lack of dedicated support \& maintenance & P15, P2, P6 \\
\midrule
\textbf{Organizational Factors} & \\[-2pt]
\textit{Regulatory Suitability (Cross-factor)} & \textit{1 subject} \\
1. Not suited for regulated organizations — \textit{\textbf{M \& O}} & P3 \\
\midrule
\textbf{Macroenvironmental Factors}  & \\[-2pt]
\textit{Community Maturity/Support} & \textit{1 subject} \\
1. Limited community support & P15 \\
\bottomrule
\end{tabular}
\label{tab:rq1_factor_view}
\end{table}

}

\paragraph{Macroenvironmental \& Organizational}
The primary weakness along this dimension related to \sigs use in large enterprise contexts.

\myparagraph{Enterprise Adoption Limitations}
Many of our participants mentioned that \sig is not suited for large enterprise applications. The issues reported by participants include rate limiting—where \sig restricts the number of signatures that can be created per unit time—along with latency concerns, where the service's turnaround time for a large volume of signatures is problematic. Additionally, participants highlighted the lack of dedicated support and maintenance teams due to \sigs open-source nature, as well as its unsuitability for organizations in regulated sectors for the same reason.
\ParticipantQuote{\Subjectseven}{We were relying on the public Sigstore instance, and I don't think it could meet our signing needs in terms of just the capacity of signatures we needed as an enterprise [rate limit].} % We would constantly be getting rate limited by the service.}
\ParticipantQuote{\Subjectthree}{For customers that are very large in scale, the biggest of the Fortune 100, or customers operating in very highly regulated environments, I think operating your own Sigstore instance within that air-gap in a private environment could be valuable. And I think it depends on the amount of software you're producing and the frequency with which you're producing it.}
% The issues here were a mixture of all usability factors\KC{this sounds off?}.

\paragraph{Technological Factors.}
We identified seven technological issues.

\myparagraph{Transparency Log Issues}
% WHile the transparency log bundled with \sig was a major advantage noted by participants, it was also a major concerns for participants. Their concerns are rooted in the public nature of the log -- where sensitive artifact metadata for company artifacts may end up in a public log of info. Another common concern was the use of the log in air-gap situations-- (what is air gap), and also the required efforts to keep manually monitoring the logs.
While the transparency log bundled with \sig was highlighted as a major advantage by participants, it was also a significant concern. Their concerns stemmed from the public nature of the log, where sensitive artifact metadata from company assets could be exposed in a publicly accessible record. Another common issue raised was the log's usability in air-gapped~\cite{nist_air_gap} environments.
Additionally, participants noted the manual effort required to continuously monitor and manage the log.
Most of these participants were in the process of experimenting with setting up the transparency log at the time of the interviews.
% A preliminary search on \sigs official documentation

\ParticipantQuote{\Subjectsix}{We've got some teams piloting Rekor, for instance, for different traceability usages...I'm all for the transparency log, but we...have to be careful who it's transparent to at what [time].}

\myparagraph{Setting up Private Sigstore Instance}
\sigs customizability was a feature frequently utilized by participants.
However, it was also associated with certain challenges.
The primary issues identified included limited documentation on setup, scarce community usage information, and the infrastructure requirements for private deployments.
Issues here were primarily related to participants' need for support resources, which were macroenvironmental and social in nature.
\ParticipantQuote{\Subjectfifteen}{There's not a big enough community of practitioners, or help to...deploy Fulcio, like on premises.} %, or in their own infrastructure.}

\myparagraph{Other Documentation Issues}
% Generally, the documentation of \sig  and practical use cases usage information and support were insufficient. Participants mostly said that the changes to the documentation were slow compared to the changes in the \sig product, clarification of the use of the different \sig components. 
Overall, the documentation for \sig, along with practical usage information and support, was found to be insufficient. Participants noted that updates to the documentation lagged behind changes in the \sig product and that there was a lack of clarity regarding the use of different \sig components.
\emph{\Subjectseventeen} highlights this, \myinlinequote{I would say the documentation lacks a bit. Just I think if you look at the documentation, it was updated quite a long time ago, like the ReadMe file. And so if we could have these different options saying that, okay, if you are starting over, this is how you can do rekor and Fulcio.}

\myparagraph{Integration to Other Systems}
% Partiicpants also wanted a wider range of compatible technologies in addition to what \siig currently offers.  Commonly mentioned integrations includes other CI/CD platforms like jenkins, Gitlab, Attestation storage databases. These unsupported technologies are highlighted to be down to \sig being  answer technology
% as highlighted by \emph{\Subjectfourteen}, \myinlinequote{}
Participants also expressed a desire for a wider range of compatible technologies beyond what \sig currently supports. Commonly mentioned integrations included other CI/CD platforms such as Jenkins and GitLab, as well as attestation storage databases. These limitations were often attributed to \sig being an emerging technology, as highlighted by \emph{\Subjectsixteen}, \myinlinequote{The weakness is not everything supports it. There's a lot of weaknesses around...its being a newer technology.}

\myparagraph{Fulcio Issues}
Some \sig issues were also tied to its certificate authority's (Fulcio) time stamping capabilities, and implementation of OIDC for keyless signing. Fulcio acts as an intermediary using OIDC identities to bind to a short-lived certificate, this was criticized by P3. P3 also commented on the lack of timestamping support in early versions of Fulcio.
\ParticipantQuote{\Subjectthree}{We do have some concerns about the way Fulcio operates as its own certificate authority. So we've been looking at things like OpenPubkey as something that removes that intermediary and would allow you to do identity-based signing directly against the OIDC provider.
And then I think the other big thing that we did, because Sigstore originally didn't support it, was time-stamping. %Because we're using Fulcio,
We want to ensure that the signature was created when the certificate was valid.}
%And so we use external time-stamping authority in order to be able to do that.}

\myparagraph{Offline Capabilities}
% Another reported \sig limitation is the absence of offline verification capabilities, via the use of \textit{offline keys}.
% This is related to the issue with the transparency log not functional in air gap environments. However the issue here is that participants wanted a capability of using \sigs signing and verification ability in an offline situations.
Another reported limitation of \sig is the absence of offline verification capabilities using \textit{offline keys}. This issue is related to the transparency log's inability to function in air-gapped environments. However, in this case, participants specifically desired the ability to use Sigstore’s signing and verification features in offline scenarios.

\myparagraph{Software Libraries}
% The programming libraries of \sig was also mentioned as being a problem, especially in situations where one wants to bundle \signs capabilities in an application.
\iffalse
The programming libraries of \sig were also mentioned as a challenge, particularly in cases where users wanted to integrate Sigstore’s signing capabilities directly into an application.
\fi
The programming libraries of \sig were also reported as a challenge, particularly when participants sought to embed signing capabilities directly into applications.
They noted the absence of well-supported, easy-to-use client libraries, which forced them to rely on workarounds such as shelling out to the CLI.

\subsubsection{Impact of Issues on Sigstore Component Use}

We assessed the effect of participants' reported problems with Sigstore on their usage of each component. 
By component, nine subjects use Cosign, nine use the Fulcio certificate authority, and nine use the OIDC keyless signing and identity manager.
Eight use the Rekor transparency log.
Only four use Gitsign and only two make use of customized components and local Sigstore deployments.
%\cref{tab: component_usage} summarizes our subjects' reported usage of Sigstore components and functionalities.

% Major components and functionalities, such as Cosign, Fulcio, and keyless signing, saw relatively high adoption. Conversely, functionalities like local deployment of Sigstore instances, which had a high number of participant-identified issues, had notably low usage. Some participants mentioned difficulties—primarily related to documentation and usage information—when setting up private Sigstore instances (\cref{tab:rq1} and \cref{sec: sigstore_weakness}
% ), consequently only two reported actually using this capability likely due to the setup difficulties.
Among these, our data shed the most light on local deployments and the Rekor log.
%Key components like Cosign, Fulcio, and keyless signing saw high adoption, while local Sigstore deployment had low usage due to numerous reported issues.
For local deployment, several participants cited difficulties with documentation and setup.
Only two successfully deployed private instances, likely due to these challenges (\cref{tab:rq1_factor_view}, \cref{sec: sigstore_weakness}).
The Rekor transparency log presents a mixed case, with both high adoption and significant reported issues.
Of the eight participants who used it, three were still in the pilot phase.
Although Rekor was recognized as a key factor driving adoption due to its strengths, it also faced a high number of reported issues, indicating that, despite its benefits, challenges persist for some participants.
% For Gitsign, while only four participants reported using Sigstore’s Gitsign, we infer that default Git commit signing was more commonly used in this context. This conclusion aligns with insights from Kalu \etal~\cite{Kalu_Singla_Okafor_Torres-Arias_Davis_2025}, who found that commit signing was widely adopted among participants.\ST{I'm not entirely sure I follow this point. People don't use gitsign, but we infer is because they are using regular git signing? I suggest this is re-phrased} 

% The Rekor transparency log presents a mixeed case of high problems reported and high adoption. Of the eight participants who reported using it, three were still piloting the component. While Rekor was cited as a key factor in adoption due to its strengths, it also had a high number of reported issues, indicating that despite its advantages, its challenges remain a concern for some participants.
\begin{figure*}[ht]
  \centering
  \includegraphics[width=0.49\textwidth]{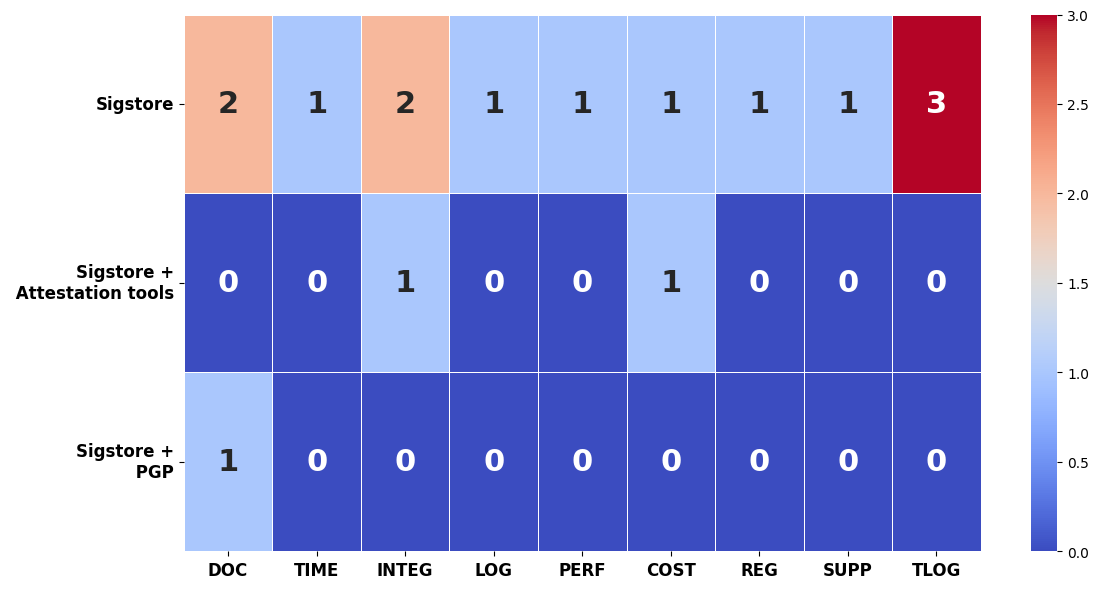}%
  \hfill
  \includegraphics[width=0.49\textwidth]{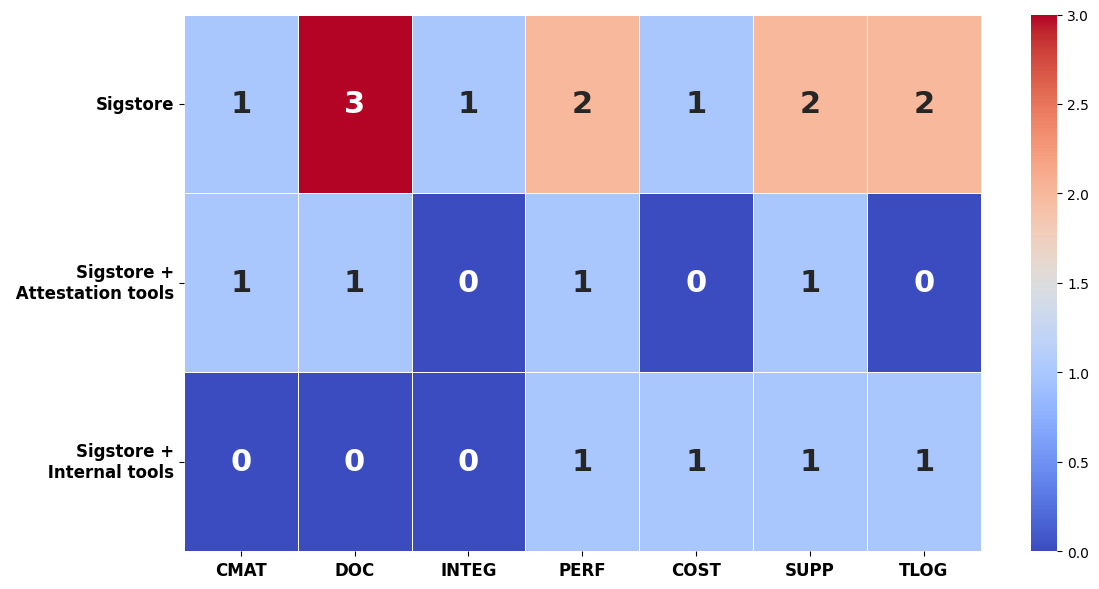}%
  \\[0.5ex] % vertical space between images and labels
  \makebox[0.49\textwidth][c]{\small \hspace{2cm} \textit{(a) Subjects from small organizations (6 subjects).}}%
  \hfill
  \makebox[0.49\textwidth][c]{\small \hspace{2cm} \textit{(b) Subjects from large organizations (7 subjects).}}%
  \vspace{5pt}
  \caption{Tools vs. Evaluation Criteria. Reported usability issues across organizational contexts.  
  Heatmaps show the number of unique participants raising each issue; darker cells indicate higher concentrations of reports. 
  \textbf{Abbreviations:} DOC = Documentation (technical); TIME = Fulcio / Timestamping Workflow; INTEG = Integrations \& Tooling; LOG = Log Monitoring Burden; PERF = Performance \& Rate Limits; COST = Private Instance: Infrastructure \& Cost; REG = Regulatory Suitability (Cross-factor); SUPP = Support \& Maintenance; TLOG = Transparency Log Constraints; CMAT = Community Maturity/Support.}
  \label{fig:heatmapsRQ1}
\end{figure*}

\paragraph{Other Components Without Mention.}
Although participants shared experiences with primary signing workflow components, discussion of other ecosystem elements was notably absent.
In particular, there was very little mention of log witnessing~\cite{hicks2023sok} or monitoring~\cite{ferraiuolo2022policy_new} for either Fulcio or Rekor.
% This is remarkable, as transparency solutions require the existence of these parties to ensure log integrity and identify log misbehavior. 
This is notable because transparency solutions rely on these parties to help ensure log integrity and detect log misbehavior.
The lack of discussion may suggest that practitioners are either not yet engaging with these components or may not fully recognize their importance in the correct use of the ecosystem.
% We believe that targeted future work should identify possible gaps in user perception, as well as identify user stories so users verify log witnesses and monitor information.
We believe future work should examine gaps in user understanding and develop concrete user stories and guidance to help practitioners verify witnesses and use monitoring information effectively.

% Although participants shared experiences with primary signing workflow components, discussion of other ecosystem elements was notably absent. In particular, there was very little mention of log witnessing [47] or monitoring [2] for either Fulcio or Rekor. 
% This is notable because transparency solutions rely on these parties to help ensure log integrity and detect log misbehavior. This absence may suggest that practitioners are not yet engaging with these components or do not fully recognize their importance for correct ecosystem use. We believe future work should examine gaps in user understanding and develop concrete user stories and guidance to help practitioners verify witnesses and use monitoring information effectively.
% \subsubsection{}

\subsubsection{Effects of Organization and Signing Tools on Identified Usability Concerns}

To provide additional context for the usability issues identified above, we highlight three observations from our analysis of organizational size and participant roles.
\ifARXIV See~\cref{fig:heatmapsRQ1,fig: appendix-HeatmapMediumfig} for visualizations.
\else
See~\cref{fig:heatmapsRQ1} for visualizations.
\fi

%Heatmaps for small and large organizations  are in.
%Medium is shown in~\cref{fig: appendix-HeatmapMediumfig}.
% (see small vs large organizational usability heatmaps -- \cref{fig:heatmapsRQ1}).

\myparagraph{Large and small organizations share similar pain points.}
Across both large and small organizations, the densest cells occur for topics related to \emph{private/air-gapped deployments}, \emph{transparency-log constraints}, and \emph{documentation for customized/ private instances}. In large organizations, technical leaders frequently surface additional concerns around \emph{support and maintenance expectations}. In contrast, smaller organizations are more willing to combine Sigstore with other tooling to compensate for gaps. Notably, large organizations in our sample were more likely to retain internal/proprietary solutions or delay full Sigstore adoption. A usability lesson here is to prioritize private-deployment guidance (including monitoring playbooks), clarify support boundaries, and provide migration patterns and user stories for mixed stacks.

\noindent
\myparagraph{Scale effects: rate limits and latency concentrate in larger orgs}
Performance and rate-limit issues are most prominent in medium/large organizations, consistent with higher CI/CD concurrency. This suggests the need for prescriptive CI recipes (e.g., batching, caching, retries/backoff, and concurrency guidance) alongside clearer Fulcio/OIDC and timestamping failure-mode documentation.

\myparagraph{Medium organizations tend to ``work around'' with plain Sigstore}
Participants from medium-sized organizations more often stick with plain Sigstore and work around gaps rather than self-host. This group benefits most from concise integration blueprints (GitHub/GitLab/Jenkins), ready-to-use attestation storage patterns, and targeted documentation updates, rather than heavy private-infrastructure guidance.

\subsection{RQ2: Why Practitioners Change Tools} %Software Signing Tools}
\label{sec:Results-RQ2}

\iffalse
We also tackle the usability of \signs tool especially as it relates to decisions that prompts the switch of practitioners in using different tools. In the context of our study, we asked participants what factors \textit{prior to adopting Sigstore} influence practitioner's to switch to \sig. The factors here are mostly Macroenvironmental (relating to regulations, user communities), Human (Previous experiences), and Technological (available capabilities of sigstore) 
\fi

% We also tackle the usability of \signs tool especially as it relates to decisions that prompts or prevents the switch of practitioners in using different tools.
% We also examine the usability of signing tools with respect to their experience and use of other tools particularly in relation to the factors that prompt or prevent practitioners from switching between different tools.

%We also examine signing‐tool usability in the context of practitioners’ experiences with alternative tools, focusing on the factors that prompt or prevent them from switching between solutions.

% \subsubsection{Current Sigstore Users -- Factors Prompting Switch to Sigstore}
% First, we asked participants who used sigstore what factors \textit{prior to adopting Sigstore} influence their (or organization) to switch to \sig. The factors here are mostly Macroenvironmental (relating to regulations, user communities), Human (Previous experiences), and Technological (available capabilities of sigstore). 
% \subsubsection{Factors Driving Practitioners to Switch to Sigstore}
\subsubsection{Drivers of Adoption Among \sig Users}
\label{sec:drivers_current_users}
\iffalse
For participants who have adopted sigstore, we enquired which factors, \emph{prior to adopting Sigstore}, influenced their decision (or their organization’s decision) to switch. 
The responses predominantly fell into three categories: macroenvironmental factors (e.g., regulations and user communities), human/social factors (e.g., prior experiences with signing tools), and technological factors (e.g., Sigstore’s unique capabilities).
\fi
For participants who adopted Sigstore, we asked which factors—prior to adoption—influenced their decision (or their organization’s decision) to switch. Their responses fell into three main categories: macroenvironmental factors (\eg regulations and user communities), human/social factors (\eg prior experience with signing tools), and technological factors (\eg Sigstore’s unique capabilities).

\JD{Need to add the paragraphs for the Cresswell factors to structure this section. From the Table, that seems pretty do-able.}

% The reasons why most of our participants chose to switch to \sig stemmed from a combination of two key factors; factors related to a subpar user experience with previous tools and factors, and factors related with \sig.
% We summarize these factors in ~\cref{tab:rq4}.
Most of our participants chose to switch to \sig due to poor experiences with previous tools and due to perceived advantages associated with \sig.  
\ifARXIV
We summarize these factors in~\cref{tab:rq2}.
\else
See~\cref{sec:appendix-TechnicalReport} for the full summary of factors.
\fi

\myparagraph{Contributions to Sigstore}
% Being a part of the ssigstore community as \textit{contributors} was a major reason for adopting sigstore. While some participants \ex P6, opined that their organization always makes a case of duty to contribute to open source communities they wish to adopt in their organization. Some of the other opinions on this was that the relative age of these teams (organizations) were relatively newer than the \sig technology i.e sigstore has been in existence prior to these organizations and the eventual members of these organizations contributed to sigstore prior to the organizational setup. Others noted that some members of their organizations played key roles in the initial setup of the sigstore project.
Being part of the Sigstore community as \textit{contributors} was a major reason for adopting Sigstore. Some participants (\eg P6) stated that their organizations have a policy of contributing to open-source communities they intend to adopt. Others noted that their organizations were relatively new compared to Sigstore—meaning that Sigstore had already existed before these organizations were established, and some of their members had contributed to Sigstore before joining their respective organizations. Additionally, some participants mentioned that key members of their organizations were directly involved in the initial development of the Sigstore project.
\emph{\Subjectfourteen} reflects this, \myinlinequote{I'd say many of the early staff, the first 10 to 20 staff were involved in the Sigstore community. There's not just our interest as a company in a Sigstore, but there's a lot of personal ties to Sigstore.} 
While this pattern may partly reflect our interview sample, it highlights how prior involvement and community ties can strongly shape adoption decisions—a dynamic that could also apply more broadly in open-source ecosystems.
% \SC{Do we think this is just because of people interviewed, or is this applicable to population}\KC{added}

\myparagraph{Available Sigstore Functionalities}
% \sigs advertised functionalities also played a role in its adoption. While a number of participants were swayed by the presence of a transparency log (which was also highlighted in \cref{sec: sigstore_strength}), its integrability to other technologies that participants were using at the time also plays a role in this decision.
\sigs functionalities also influenced its adoption. Several participants were drawn to the transparency log (as highlighted in \cref{sec: sigstore_strength}).
\ParticipantQuote{\Subjectfive}{Sigstore has other features, like you can use your OIDC identities, like your GitHub identities as well, to do the signing, [and] they also maintain a transparency log.}

Integration with existing technologies also affected their decision.
As \emph{\Subjectone} said, \myinlinequote{Factors I considered were, ..., the existence of tools to use it.}

\myparagraph{Regulations and Standards}
% The impact of regulations and standards in adopting \sig was also pronounced amongst our participants. Some of our participants whose organizations are affected by specific security regulations on signing and attestations find that sigstore's integrability with attestations helped with those requirements. Also, some organizations while trying to be more secure, some of our participants note that some security frameworks/standards recommend\sig.
% The impact of regulations and standards on adopting \sig was also significant among our participants. Some participants who were using other tools (2), were moved by these regulations to adopt toward sigstore since that certain security frameworks and standards recommend \sig as part of their best practices. Additionally, some participants do note that Regulations played somewhat of a role in their adoption of \sig.
The impact of regulations and standards on adopting \sig was significant among our participants. Some participants who previously used other tools (2) switched to \sig due to security frameworks and standards recommending it as a best practice. Additionally, the remaining participants (2) noted that regulations (\eg EO-14028) played a role in their decision to adopt \sig.

\ParticipantQuote{\Subjectfive}{...standards actually really helped to convince everyone that we should be doing this ...
But once a standard is put in place, once the requirements are set, then it just sped up the process.} % of adopting these new technologies like Sigstore ...}

\myparagraph{User Community \& Trust Of Sigstore Creators}
% The large user community of sigstore was also a key participant reason for adopting \sig.
% This corresponds totother reason by some participants(2 -- P1, P10) who complained of GPG's small user community. 
% Another similar factor is the inherent trust of the developing community of CNCF (Cloud Native Computing) security tolings which the participant have grown to trust over time.
\iffalse
The large user community of \sig also motivated its adoption among participants.
Conversely, participants P1 and P14 cited GPG's smaller user base as a drawback.
Additionally, some participants expressed trust in industry consortia such as CNCF (Cloud Native Computing Foundation) 
\fi
The large user community of \sig also motivated its adoption among participants. Conversely, participants P1 and P14 cited GPG’s smaller user base as a drawback. In addition, some participants expressed trust in industry consortia such as the CNCF (Cloud Native Computing Foundation), noting that Sigstore’s development and governance under CNCF gave it additional legitimacy and credibility compared to tools maintained by smaller, independent groups.
 % security tooling community, which they had grown to rely on over time.
\ParticipantQuote{\Subjectthree}{We do defer often at the higher-level projects to different foundations. So we would be more likely to trust something from the CNCF...[than] independent developer projects.}
%or Sigstore rather than small little independent developer projects or things like that.}

\myparagraph{GPG \& Notary Issues}
% The issues associated with the prior existing tools  played a major role in shaping the decisions of partitioners in switching to sigstore. GPG based signing implementations have being historically criticised for key management issues, and ease of user, and our participants mentioned these issues in addition to the low adoption rates, learning curves and integrability with modern technology.
% Notary issues were mostly around customer demands, compatibility with other technologies, and non-regular updates. Key and Identity  management issues was common to both GPG and Notary. Users of proprietary signing tooling are plagued difficult setup.
\iffalse
Issues associated with previously existing tools played a major role in practitioners' decisions to switch to Sigstore. GPG-based signing implementations have historically been criticized for key management challenges and usability concerns. Our participants echoed these issues, citing additional factors such as low adoption rates, steep learning curves, and limited integration with modern technologies.

Notary-related issues primarily concerned customer demands, compatibility with other technologies, and infrequent updates. Key and identity management challenges were common to both GPG and Notary. Meanwhile, users of proprietary signing tools faced difficulties with complex setup processes.
\fi
Issues with previously existing tools played a major role in practitioners’ decisions to switch to Sigstore. Both GPG- and Notary-based signing implementations were reported as problematic. GPG has long been criticized for key management challenges and usability concerns, and our participants echoed these issues, citing additional factors such as low adoption rates, steep learning curves, and limited integration with modern technologies. Notary-related issues primarily concerned customer demands, compatibility with other technologies, and infrequent updates. Key and identity management challenges were common to both GPG and Notary. Meanwhile, users of proprietary signing tools faced difficulties with complex setup processes.
% \SC{should probably bring up Notary in the first paragraph}\KC{better?}

%\subsubsection{Barriers and Motivators Spurring Consideration of Sigstore Among Non-Sigstore Users}
\subsubsection{Considerations from Non-\sig Users}
\label{sec:drivers_non_users}

In this section, we highlight reasons mentioned by practitioners that have stopped them (or their organization) from switching to identity-based tools like Sigstore (see \cref{tab:non_sigstore_reasons_against}). We also report issues experienced by these non-Sigstore users that have prompted a consideration (or in some cases partial adoption) of these identity-based tools.

\input{misc/data/table-nonsig-2}

\myparagraph{Barriers to Considering Sigstore}
% As highlighted in \cref{tab:non_sigstore_reasons_against}, most reasons mentioned by our subjects for not adopting sigstore stems from organizational reasons.
% \ParticipantQuote{}{}
% Other notable issues stemmed from \sigs technical issues ranging from privacy concerns about using sigstore's public instance instance, and the inherent risk introduced by essence of it being a third party tool maintained by open source maintainers.
As highlighted in \cref{tab:non_sigstore_reasons_against}, most reasons for non-adoption were organizational.
\ParticipantQuote{\Subjecteight}{When it comes to open source projects and open source stuff, our organization is skeptical. Policy and everything drives teams away...There's
%So things like Sigstore would have to go through extensive security or review process and whatnot.
also the ongoing risk of having a product that's built outside continually shipping code and being integrated into this product, where it deals with sensitive material. We do evaluate these things, but the scrutiny against code written outside of the company is extreme, and with good reason given the [supply chain] context.}
Other notable issues arose from technical concerns, like privacy risks associated with Sigstore’s public instance and Macro environmental concern of the inherent risk of relying on a third-party tool maintained by open-source contributors.
\SC{i feel like issues arising from technical concerns could be worded better}

\myparagraph{Factors Motivating Consideration for Sigstore}
% \JD{I moved~\cref{tab:non_sigstore_reasons_for}, so crefs to it should be changed/removed.}
% As shown in \cref{tab:non_sigstore_reasons_for} when these practitioners were asked about issues they have encountered using their current tools and considered fixes, Mostly current tooling issues are the main driving points for any sort of considerations \eg drawing design inspirations for internal tools, partially adopting some identity-based tools.
% While the subjects who reported drawing designs from identity-based workflows they specifically don't mention sigstore
% As shown in \cref{tab:non_sigstore_reasons_for},
When practitioners were asked about issues with their current tools and potential fixes, most pointed to problems with existing tools as their main driver to alternatives, \eg incorporating identity-based features into internal tools or partially adopting new solutions. Although some subjects reported drawing design inspiration from identity-based workflows, none specifically mentioned Sigstore.
\emph{\Subjecteleven} states, \myinlinequote{I'm sure ... [the internal signing tool] was influenced by [an open-source identity-based tool]...they adopted it and made the right tweaks internally.}
Lack of transparency and integration of these tools to other platforms were also common.
\ifARXIV We summarize these factors in~\cref{tab:non_sigstore_reasons_for}. \else We summarize these factors in~\cref{sec:appendix-TechnicalReport}. \fi

% \ifEXTENDED
% We summarize these factors in~\cref{tab:non_sigstore_reasons_for}.
% \else
% % For a complete summary of issues, see ~\cref{sec:DataAvailability}.
% See~\cref{sec:DataAvailability} for full issue details.
% \fi

\subsubsection{Effects of Organizational Context} % Effects on Adoption of Signing Tools}
\label{sec:org_context}

\begin{figure}
    \centering
    \includegraphics[width=0.95\linewidth]{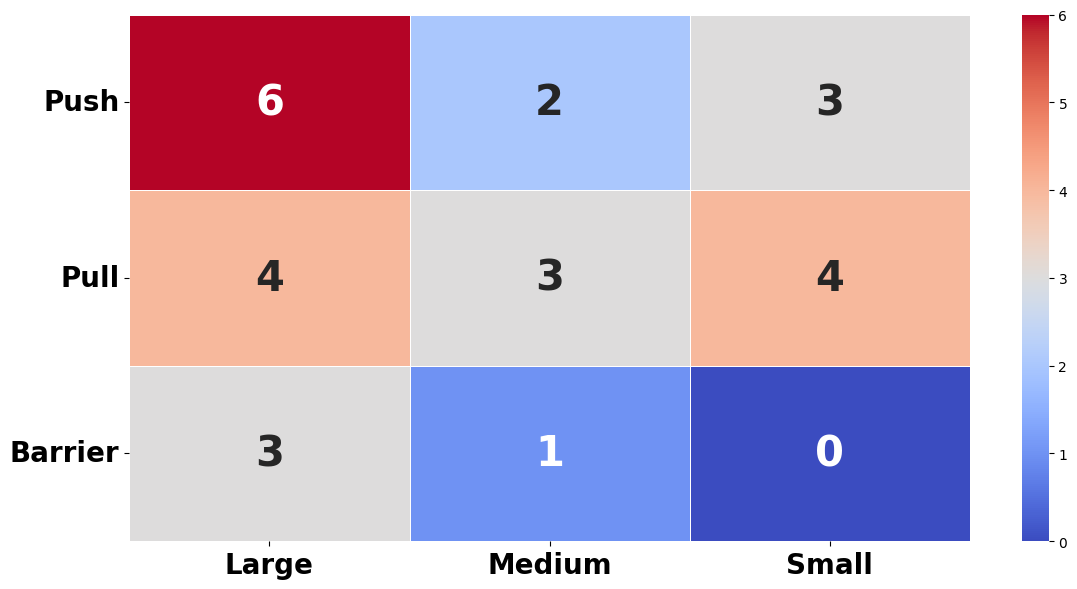}
    \caption{Push-pull-barrier heatmap.
    % \SC{0 should be white, six is also kind of hard to read}
    }
    \label{fig:PushPullHeatmap}
\end{figure}

\ifARXIV
Combining the results in \cref{sec:drivers_current_users} (\cref{tab:rq2}) and \cref{sec:drivers_non_users} (\cref{tab:non_sigstore_reasons_for,tab:non_sigstore_reasons_against}) with organizational size, former signing tools, and current signing tools reveals several patterns. We summarize these in the Push-pull-Barrier heatmap (\cref{fig:PushPullHeatmap}).
\else
Combining the results in \cref{sec:drivers_current_users}  and \cref{sec:drivers_non_users} (\cref{tab:non_sigstore_reasons_against}) with organizational size, former signing tools, and current signing tools reveals several patterns. We summarize these in the Push-pull-Barrier heatmap (\cref{fig:PushPullHeatmap}).
\fi

\paragraph{Push, Pull, and Barrier Factors Influencing Adoption.}
We grouped issues raised by participants into \emph{Push}, \emph{Pull}, and \emph{Barrier} factors. We define these factors drawing on Bansal \etals Push-Pull-Mooring framework for consumer switching behavior~\cite{bansal2005migrating}.
\emph{Push} factors are problems with prior tools that drove current users toward Sigstore.
\emph{Pull} factors are attractions of the new tool (and ecosystem). % that drew %both current users and non-users.
\emph{Barriers} are reasons that stalled or prevented adoption. 

While large organizations exhibit the greatest number of barriers, they report push and pull factors in roughly similar measure. 
\SC{it would be helpful to have the following sentences be organized as a summary of each of the factors: is a bit confusing as to how the previous and following sentences connect right now}
By current-tool status, many push factors come from teams now using Sigstore alongside other mechanisms (e.g., attestations, internal tooling, PGP), which often reflects phased introduction of Sigstore into existing processes (and vice versa). \ifARXIV A summary of the coded factors appears in \cref{tab:rq2_codes_drivers} \else A summary of the coded factors appears in \cref{sec:appendix-TechnicalReport}\fi. A usability takeaway is that switching is \emph{primarily} driven by human/social push (e.g., frustration with legacy workflows) and only \emph{secondarily} by pull from Sigstore features and regulatory pressure (Sigstore being referenced by some standards). Thus, even as technical innovations “pull” adoption, community-building and other social supports also materially “push” change.

\paragraph{Non-Adoption Reasons Are Often Organizational and Non-Technological.}
\iffalse
Although non-adopters are both pushed by shortcomings in internal tools and pulled by identity-based capabilities, adoption frequently stalls due to non-technological (predominantly organizational) factors. This suggests identity-based providers should reassess service models: several enterprises want private deployments and stronger support/SLAs. Tools originally aimed at open-source uptake may benefit from enterprise-tailored options (private, well-supported offerings) to improve adoption in larger organizations.
\fi
Non-adoption reasons are often organizational and non-technological. Although non-adopters are both pushed by shortcomings in internal tools and pulled by identity-based capabilities, adoption frequently stalls due to organizational constraints such as risk posture, support expectations, or regulatory requirements. This pattern is not unique to software signing—similar dynamics have been observed in other infrastructure domains (e.g., cloud technologies\cite{al2021systematic})—but it is particularly salient here because identity-based signing tools must be trusted across organizational boundaries. Our interpretation is therefore not that identity-based tools have failed, but rather that their technical attractiveness has created demand for more mature deployment models: private instances, enterprise-grade support, and stronger service guarantees.

\paragraph{Tool Migration Patterns.}
In our sample we observe recurring migration paths:  \emph{PGP} $\rightarrow$ Sigstore (sometimes retaining PGP during transition); \emph{Notary v1} $\rightarrow$ Sigstore or \emph{Notary v2}; and \emph{internal/proprietary} $\rightarrow$ Sigstore (or hybrid). Organizations mostly transition between tools slowly. Larger organizations are more likely to prefer internal/hybrid arrangements (reflecting policy, privacy, and support concerns), whereas small/medium organizations are more likely to adopt “pure” Sigstore sooner, using hybrids as transitional steps or to meet offline/private constraints.

\section{Discussion} 
\label{sec:Discussion}

We highlight the main takeaways from our results. 
This research is a case study of identity-based software signing tools (described in \cref{sec:Background-SoftwareSigning}) using Sigstore.
We discuss implications for this class of tools along three axes: a technological readiness assessment, their transformative potential, and the implications of the observed usability patterns.

\subsection{TRLs and Identity-Based Signing Tools}
%“Technological Readiness” of identity-based Signing Tools}
\label{sec:trl_next_gen_signing}
Because Sigstore, our exemplar of identity-based signing, is relatively recent, the maturity of its core capabilities is uneven. Understanding this maturity helps organizations assess implementation risk and prioritize investment \cite{mankins2009technology}. Our usability study provides a preliminary signal.
Drawing on participant reports and interpreting them through the Technology Readiness Level (TRL) lens \cite{mankins2009technology}, we note that TRL is a technology-maturation framework (not a usability scale). In practice, lower TRLs (concepts, prototypes) tend to surface mostly \emph{technical} concerns, while higher TRLs (operational environments) bring \emph{non-technical} constraints (organizational fit, governance, support/SLAs, regulation) to the foreground. Our results fit this progression: \emph{technological} factors were most frequently discussed. Several capabilities, \eg short-lived (ephemeral) keys, OIDC-based signer identity, and GitHub Actions workflows, received consistently positive mentions across different organizational contexts, suggesting higher maturity for these components (and analogous features in similar identity-based tools). In contrast, capabilities such as private/self-hosted deployments and operation in regulated environments surfaced moderate friction, indicating mid-level readiness and a need for stronger documentation, reference architectures, and support pathways.
For organizations that have not yet adopted identity-based tools, these findings suggest comparatively lower risk for the first set of capabilities, while private/regulated deployments warrant careful evaluation and piloting. These assessments are preliminary and should be read in light of our case-study scope.

\noindent
\myparagraph{From Sigstore to the Broader Class}
Sigstore demonstrates that identity-based signing can work at scale (across multiple organization sizes and use contexts in our sample), serving as a proof-of-concept for the class. Where we observed Sigstore-specific friction (e.g., private Fulcio/Rekor deployments, public-log privacy trade-offs), the underlying \emph{class} implications are clear: enterprise deployment models, integrations, and operational guardrails are the next maturity hurdles \emph{for identity-based signing generally}.

\iffalse
\subsection{The transformative potential of Sigstore}
When compared with other identity-based signing tools, Sigstore incorporsates and pilots several capabilities that are yet to be incoporated by other tools; example-oOpenpubkey\cite{} currently only implements the core functionality of identity-based software signni tools which ids identity management (tying OIDC identitiy to key maangement). The amerits of using Sigstore as mentioned by our participants are msotly centered on the availability of these additional capabilities like the Rekor Transpaerncy logs, compatibility with other technologies like attestaions , container registries etc even though these capabilities have some usability concerns, the ambitiousness of these innovations has been greatly leveraged by the sofwtare engineering and security rersearch communities as building blocks to pilot , research and propose improvement s that will be felt across the software supply chain security landscape (it has been adopted by maven, pypi, kubernetes etc). Ss much as these issues has been listed they have Sigstore has proviided a research base to define and propose innovations to solve tese and other emerging issues; Proivacy concerns- \cite{merrill2023speranza}, diversification of OIDC Id management capabilities \cite{okafor_diverify_2024} etc.
\fi

\subsection{Transformative Potential}

% Sigstore, an exemplar of identity-based (identity-based) signing, bundles capabilities beyond core identity-to-key binding (e.g., OpenPubKey focuses primarily on OIDC–key binding). In addition to OIDC-based signer identity and short-lived keys, Sigstore provides a transparency log (Rekor) and broad ecosystem compatibility (attestations, container registries, CI/CD). Participants repeatedly cited these \emph{additional} capabilities as differentiators—even while noting usability frictions. The ambition of this feature set has been leveraged by the software engineering and security research communities as building blocks to pilot, study, and propose improvements with ecosystem-scale impact (e.g., adoption or integration by Maven, PyPI, Kubernetes). In parallel, the project has catalyzed research directions such as privacy-preserving transparency~\cite{merrill2023speranza} and diversified OIDC identity provisioning~\cite{okafor2024diverify}.

Sigstore, as the most widely deployed exemplar of identity-based signing, goes well beyond the minimal OIDC–key binding model offered by alternatives like OpenPubKey. In addition to identity-based signing and short-lived keys, Sigstore integrates a transparency log (Rekor) and broad ecosystem compatibility with attestations, container registries, and CI/CD workflows. Participants repeatedly emphasized these \emph{additional} capabilities as unique differentiators, even while reporting usability frictions.

This dual reality—strong adoption coupled with persistent friction—underscores Sigstore’s role as a living testbed for both practice and research.
\sig is already being leveraged in production ecosystems (\eg Maven, PyPI, Kubernetes) and has catalyzed new research trajectories, such as privacy-preserving transparency~\cite{merrill2023speranza} and diversified identity provisioning~\cite{okafor2024diverify}.
The broader implication is that Sigstore is not merely a tool to be adopted, but an infrastructure on which future signing solutions and empirical studies can (and are) being built. Widespread adoption makes its usability challenges consequential --- and resolving them has cascading benefits across the supply chain. Its design decisions and shortcomings provide researchers with concrete leverage points for advancing the next generation of signing tools.

\subsection{Usability Lessons for ID-Based Signing}

\iffalse
Our results reveal different adoption drivers—and shared usability concerns—between Sigstore users and non\mbox{-}users. Sigstore adopters most often cited contributions to the Sigstore project (a human/social factor) and macroenvironmental pressures (e.g., standards), alongside negative experiences with legacy tools, as primary motivators. In contrast, non\mbox{-}users emphasized organizational considerations when choosing their signing solution.

Despite these differences, both groups reported similar technological usability issues, particularly around integration with other systems and infrastructure transparency. Notably, non\mbox{-}users frequently highlighted signer identification challenges, whereas Sigstore users did not—underscoring a clear usability advantage of identity\mbox{-}based tooling for identity management.
\fi
\iffalse
Our results reveal different adoption drivers—and shared usability concerns—between Sigstore users and non\mbox{-}users\cref{sec:Results-RQ2}.
Both groups reported some similar technological usability issues, particularly around integration with other systems and infrastructure transparency of these sigingn setups.
This suggest a cross sutting concerns that affects not just sigstore users but also non sigstore users. A preliminary investigation for some other identity-based signing tools like open-pubkey shows simialr (from the enhancement issue logs\cite{}). We distill actionable lesons for identity-based tools adn opportunities for future works.
\fi
While our results reveal different adoption drivers between Sigstore users and non\mbox{-}users, there exists some shared usability concerns—between Sigstore users and non\mbox{-}users (\cref{sec:Results-RQ2}). Both groups reported similar technological issues, particularly around integration with other systems and the transparency of signing infrastructures. This suggests cross\mbox{-}cutting concerns that affect not only Sigstore users but also non\mbox{-}users. A preliminary look at enhancement/issue trackers for other identity\mbox{-}based signing tools (\eg OpenPubkey) shows similar themes~\cite{OpenPubKey_Issues}. Building on these observations, we distill actionable lessons for identity\mbox{-}based tools and outline opportunities for future work.

\myparagraph{Automated Cross\mbox{-}Platform Integrations}
To fully leverage flexibility, practitioners need maintained integration modules for popular CI/CD systems, registries, and orchestration platforms. Missing integrations and sparse documentation force teams to build custom connectors, increasing misconfiguration risk and undermining security guarantees.

\myparagraph{Need for Inter\mbox{-}Platform Standards}
Standardization of information exchange is a typical problem for new infrastructure (\eg SBOM~\cite{Xia_Bi_Xing_Lu_Zhu_2023}). Identity\mbox{-}based signing would benefit from uniform schemas for transparency events and identity assertions.
Recent efforts like DiVerify~\cite{okafor2024diverify}, which harmonizes identity provisioning across multiple OIDC providers, take a step toward smoother data flows between signing components, transparency services, and external toolchains.

\myparagraph{Verification Workflow}
A core supply\mbox{-}chain property is transparency of actors and artifacts, which requires reliable, usable verification of signer identities~\cite{okafor_sok}. Prior work reports verification pain in both organizational and open\mbox{-}source settings~\cite{usenix_2025_signing_interview_kalu, schorlemmer2025establishing}. Although Sigstore provides multiple verification paths (Cosign CLI, transparency log), participants reported log\mbox{-}related usability issues and limited automation elsewhere. Further automating verification would better realize the security benefits of this tool class.

\myparagraph{Privacy Concerns}
Privacy concerns emerged prominently around Sigstore’s Rekor transparency log: five of the six participants who reported concerns with Rekor cited privacy as a significant issue. Similar concerns also surfaced for keyless signing via OIDC. These are not unique to Sigstore~\cite{merrill2023speranza} but reflect broader tensions in cryptography between transparency and confidentiality. This underscores the need for privacy\mbox{-}preserving designs (e.g., selective disclosure, redaction regimes, witness policies) that balance auditability with confidentiality. In Creswell’s terms, these concerns span macroenvironmental (policy/regulation), organizational (data governance and risk posture), and human/social (developer risk perception) factors, reinforcing that privacy is not solely a technical hurdle but a cross\mbox{-}factor adoption barrier.

\myparagraph{Usability as a Limit on Realized Security}
For security tools, usability problems are barriers to realizing effective security\cite{green_developers_2016}. When tools are hard to use, their features may be partially deployed, misconfigured, or disabled\cite{green_developers_2016}. Our findings show this in the context of identity-based signing: for example, rekor transparency-log usability concerns mean that some corporate users avoid the log feature, despite Sigstore supporting more privacy-preserving setups (which seem to be unusable in their context).
As a result, these users miss the full auditability the log can provide.
For example, \emph{\Subjectthree} said, \myinlinequote{Well, I think there's definitely reasons we haven't adopted some things like Rekor, and the biggest reason for that is being able to validate everything offline in our capped(air-gapped) environments. And so there's not a whole lot of value for us in having something in the transparency log if we can't prove somewhere
else without access to Rekor, that it's in the transparency log.}

\myparagraph{Population Scope \& Future Populations}
Our study primarily sampled experienced practitioners operating within organizations, which shapes both the opportunities and frictions we observed. Usability needs likely differ for other communities and skill levels—\eg individual open-source maintainers, small/startup teams, and security novices. %engineers with limited security background or constrained resources.
In light of the biases from our particular sample, we recommend further research that
  (1) uses organizational contexts and user personas as a primary selection criterion;
  and
  (2) considers distinct signing use cases for a more fine-grained understanding.

\subsection{Contrast to Prior Work}

\myparagraph{Prior Work on Adjacent Secure Communication Tools}
We restate that most prior work performs in-vitro usability experiments with novices using general-purpose encryption tools (email, messaging clients). In contrast, we studied expert practitioners in-vivo focused on identity-based software signing for software supply chain security. We observe some similarities in usability factors, eg integration with existing technologies \cite{atwater_leading_2015} and trust in the tool\cite{atwater_leading_2015, ruoti_confused_2013, fahl_helping_2012}. We also observe differences: our study design gives us insight into organizational and regulatory constraints, and tool adoption/migration pathways (\cref{sec:org_context}, \cref{sec:trl_next_gen_signing}).

\myparagraph{Generalizability to Other Software Engineering Tools}
Beyond adjacent secure communication tools, our findings also speak to other classes of software engineering tools.
Our findings reinforce several adoption factors noted in prior literature on open source tools\cite{del_organizational_2010}, such as ease of use, compatibility with existing technologies, and integration flexibility, which were especially salient for identity-based signing tools like Sigstore. They also highlight additional considerations that may generalize to other classes of secure software engineering tools example the role played by macroenvironmental factors like the tooling community, and regulations\cite{Lenarduzzi_open_2020}.

Furthermore, our results suggest adoption patterns not widely discussed in existing research. In particular, we observe the emergence of a migration pathway model, where teams incrementally adopt identity-based tools. Several non-Sigstore users reported experimenting with or partially integrating smaller solutions, such as OpenPubKey and TUF, to meet specific needs like key delivery and management. This suggests a stepwise transition process in tooling adoption that may be common across emerging tool ecosystems. This model can also be important to understand and further propose a streamlined organizational adoption pathway for software engineering tools.

% Our findings illustrate this in the context of identity-based signing: for example, transparency-log usability and privacy concerns lead some corporate users to avoid the log feature, even though Sigstore supports more privacy-preserving configurations. As a result, these organizations do not obtain the full auditability and compromise-detection benefits the log is designed to provide, \eg \myinlinequote{p3}{P3: "Well, I think there's definitely reasons we haven't adopted some things like Rekor. And the biggest
% reason for that is being able to validate everything offline in our capped environments. And so there's
% not a whole lot of value for us in having something in the transparency log if we can't prove somewhere
% else without access to Rekor that it's in the transparency log}

\section{Threats to Validity}
\label{sec:Threats}

% \JD{Hmm, be clearer about what we are taking from Verdecchia \etal (they said to focus on the real threats not just fluff)}
% We discuss the internal and external threats to the validity of our works following Verdecchia \etals~\cite{verdecchia2023threats} recommendations.
% We additionally describe the role our experiences and perspectives may have played in our research
To discuss the limitations of our work, we follow Verdecchia \etal's \cite{verdecchia2023threats} recommendation to reflect on the greatest threats to our research. %, identify mitigations, and discuss the applicability of our work.
In addition, we discuss how our perspectives may have influenced our research (positionality).

\myparagraph{Construct Validity}
%Construct validity is concerned with how accurately our tests and chosen frameworks accurately measure our results.
% Cresswell's usability framework was developed for health information technology.
% We applied it to software signing tooling, but may have introduced flaws.
% Cresswell’s framework was designed for health IT, where “tooling” refers to integrated clinical systems. Applying it to software signing poses construct risks. However, the core factors (Technological, Social/Human, Organizational, Macroenvironmental) remain relevant. We mitigated this risk by inductively validating each theme against practitioner interviews to ensure our adapted constructs accurately reflect signing-tool usability.
Cresswell’s framework targets health IT, where “tooling” denotes integrated clinical systems, so applying it to software signing carries construct risks. 
% However, its core factors (Technological, Social/Human, Organizational, Macroenvironmental) remain relevant. 
We mitigated risks by inductively validating themes (from its core factors) against practitioner interviews, ensuring our adapted constructs reflect signing-tool usability.

%\JD{External Validity? Construct Validity (definition and interpretation of usability)?}
\myparagraph{Internal Validity} Threats to internal validity reduce the reliability of conclusions of cause-and-effect relationships. An additional internal threat exists in the potential subjectivity of the development of our codebook. To mitigate this subjectivity, we utilized multiple raters and measured their agreement.
% A significant portion of our participants reported issues related to inadequate documentation and lack of practical use cases. Notably, our interviews were conducted between November 1, 2023, and February 9, 2024. We also observed that the \sig documentation (and components) ~\cite{sigstoreDocs} underwent significant updates later in the year. Thus while we anticipate some of these problems may have been fixed, we still underscore the importance of our findings in helping drive other \signs tool ecosystems to improve the usability of their tools.

\myparagraph{External Validity}
The primary generalizability threat is that we recruited from a population biased toward the use of Sigstore. 
%to external validity influence the generalizability of research and applicability to further contexts. As we evaluate how a signer's decision to use a specific tool changes over time (RQ2), a confounding variable that may impact external validity stems from the experienced technical background of our participants, all of whom are already using Sigstore.
%Our results may not reflect the perspectives of practitioners who are considering switching to Sigstore, or those who have transitioned to other new toolings that are not Sigstore. 
We viewed this as necessary in order to recruit sufficient subjects.
%The result, a focused treatment of usability in Sigstore, comprises a valuable case study. %contribution to the literature.
%This is implicit in our Our research serves as a targeted case study of the Sigstore ecosystem and does not seek to make generalizations and conclusions about other toolings.
A second threat arises from the sample size of our work, which is due to the challenges of recruiting expert subjects.
%cost constraints associated with the costs of compensating interview study participants. Furthermore, some participants of our interview study were associated with a Linux Foundation-hosted conference, which can introduce a positive bias towards Sigstore as a tool.
%To mitigate this, we achieved code saturation, leading us to conclude that no new themes would have emerged with additional data from a larger sample.
%temporal validity.
%For Many our participants reported issues related to inadequate documentation and lack of practical use cases. Of note,
\iffalse
A third threat is temporal: our interviews were conducted between November 1, 2023, and February 9, 2024, and reflect the experiences of Sigstore users at that time.
Since then, \sigs updates—including new signature schemes, bug fixes, and dependency upgrades—have been entirely additive, with no changes to its core workflows or components, suggesting our findings' continued relevance beyond specific releases
\fi
Third, our study relies on an expert population, whose usability concerns may not reflect those of novice users.
Another threat is our study's temporal scope.
Interviews were conducted between November 1, 2023, and February 9, 2024, capturing Sigstore users’ experiences during that period.
Sigstore has changed since then, but not the %additive changes—such as new signature schemes, bug fixes, and dependency upgrades—without modifying
core workflows and components.
We believe our findings remain relevant.

%may not  We also observed that the \sig documentation (and components) ~\cite{sigstoreDocs} underwent significant updates later in the year. While we anticipate some of these problems may have been fixed, we still underscore the importance of our findings in helping drive other \signs tool ecosystems to improve the usability of their tools. 

\SC{Slack comment: ``Also, if we have space for discussion on further work, would be interesting to talk about the importance/need to understand usability for other types of communities and skill levels (ex individual software engineers making some open source tool vs software engineers in the context of an organization; usability for software engineers with differing amounts of security background and understanding)"}

\SC{Proposal: we focused on sigstore as used by mostly experienced professionals in organizations; since context changes between team to team, usability needs may change based on experience of professionals as well as access to resources}

\myparagraph{Statement of Positionality}
The author team has expertise in software supply chain security in general and software signing methods in particular.
They also have expertise in qualitative research methods.
This background qualifies the team to design, conduct, and analyze the data described in this study.
In addition, one of the authors is a Sigstore contributor.
Their inclusion enabled the author team to better interpret and contextualize some of the interview data.
To avoid bias, this author was not involved in developing the research design or conducting initial analysis of the results, but provided insight in later analysis.

\section{Conclusion}
\label{sec:Conclusion}

Software signing underpins supply chain security by guaranteeing provenance and integrity, yet its effectiveness depends on usable tooling. Identity-based systems like Sigstore reduce legacy key-managed burdens of key management and signer identification, but our study shows that adoption is shaped by both strengths and frictions. Identity-based workflows, ephemeral keys, and CI/CD compatibility ease use, while integration gaps, transparency log privacy concerns, and organizational constraints remain barriers—especially in large enterprises.

\iffalse
Our study shows that Sigstore is not only a valuable exemplar of identity-based signing but also a lens into broader usability issues that will shape the future of software signing. While its automation and short-lived keys help overcome long-standing barriers in legacy key-managed signing tools, our findings surface several cross-cutting lessons. Usability must extend beyond user interfaces to include organizational policy and workflow integration; transparency and privacy need careful balancing; and ecosystem maturity (documentation, integration tooling, support) is decisive for adoption. These insights matter directly to Sigstore maintainers and adopters, but they also generalize to other emerging platforms (\eg OpenPubKey, AWS Signer, SignServer) that share identity-driven design patterns. By surfacing these factors, we underscore that usability is central to supply-chain security: without attention to adoption barriers, even the best technical mechanisms will fail to achieve their security potential.
\fi

By surfacing these factors, we show that usability is central to supply-chain security.
Without attention to adoption barriers, even the best technical mechanisms will fail to achieve their security potential.
Our findings reveal uneven maturity across \sigs components, and suggest lessons for other signing toolmakers, \eg OpenPubKey and SignServer, which share similar design patterns. % with some ready for broad use and others requiring stronger documentation, support, and enterprise deployment models.
We recommend that toolmakers prioritize official integrations and configurable privacy controls to improve usability and trust.
By addressing these challenges, identity-based signing tools can expand the coverage of verifiable artifacts and attestations, shrinking supply chain attack surfaces to improve software security.
% Software signing is widely recommended to ensure software provenance, yet many organizations face challenges when integrating it into their processes. In this first qualitative study, we examine the practices and challenges of software signing through interviews with 18 senior practitioners across 13 organizations. We identified that while software products are signed, the signatures often go unvalidated. Key barriers include infrastructure limitations, lack of expertise, and insufficient resources. Although signing offers strong provenance guarantees, it remains secondary in many organizations' cybersecurity strategies. Additionally, organizational behaviors are influenced by major incidents like SolarWinds and industry regulations and standards, with concerns raised about the quality and development of these standards.

% Our findings shed light on the real-world practices, challenges, and organizational factors shaping software signing adoption, offering guidance for organizations on their journey toward more secure software supply chains.
\iffalse
\section{Data Availability}
\label{sec:DataAvailability}

An artifact containing our interview protocol, codebook, etc. is available at: \url{https://github.com/nextgenusability/identity-based-usabilty}.
Due to the specialized target population as well as our recruiting method, to ensure subject anonymity we will not share the interview transcripts.
\fi

\ifANONYMOUS
\else
\section{Acknowledgments}
We thank the study participants for contributing their time to this work.
We acknowledge support from Google, Cisco, and NSF \#2229703.
We thank Kimberly Hung for assistance with figure preparation.
We also thank Hayden Blauzvern and all reviewers who provided feedback on this work.
\fi

% \clearpage
% \raggedbottom
% \pagebreak
% % optional clearing of the page
% \cleardoublepage
\appendix
% \section*{Ethical Considerations}
% \textbf{Within up to one page, explain the ethical considerations of your work. This appendix must have exactly this title, otherwise you will risk desk rejection. Carefully study the Ethics Guidelines before submitting your paper.}
% % optional clearing of the page
% \cleardoublepage

\section*{Ethical Considerations}

Our study investigates the usability of identity-based software signing tools, focusing on Sigstore as an exemplar. We conducted semi-structured interviews with 17 security practitioners across 13 organizations, analyzing their perspectives on usability strengths, weaknesses, and adoption dynamics.

\paragraph{Stakeholders \& Potential Harms.}  
Following the guidance of Davis \etal~\cite{davis2025guide}, we identify the following stakeholders.  

\begin{itemize}[leftmargin=12pt, rightmargin=5pt]
\item \emph{Direct stakeholders:}
  (i) Interview participants, whose responses could unintentionally expose their organization’s practices;
  (ii) The organizations employing these participants, which may face reputational, regulatory, or security risks if weaknesses are revealed;
  (iii) The research team, responsible for balancing transparency and confidentiality;
  and
  (iv) Adversaries, who could misuse disclosed weaknesses in signing tools or deployment practices.  
\item \emph{Indirect stakeholders:}
  (i) Other software-producing organizations, who may be affected if the study highlights industry-wide shortcomings in the signing tool used by such organizations;
  (ii) Standards bodies, whose guidelines may be challenged;
  (iii) Software consumers, whose trust in signed artifacts could be affected;
  and
  (iv) The broader public, for whom systemic weaknesses in signing infrastructure may undermine confidence in software supply chains.
\end{itemize}

% \textbf{Potential Harms.}  
% First, practitioners risk reputational or professional harm if their identities are not fully anonymized. 
% Second, organizations may be exposed to reputational or regulatory scrutiny if weaknesses in their signing practices are revealed. More critically, adversaries could exploit weaknesses in Sigstore or similar tools if technical details of flaws or poor practices are publicized. Highlighting systemic weaknesses also risks undermining societal trust in software signing and supply chain security, even as the study’s intent is to improve adoption and resilience.

\paragraph{Mitigations Implemented.}  
\emph{Recruitment Policies:} Data were collected through voluntary interviews, with written consent obtained prior to participation. Recruitment was conducted through established professional networks; no unsolicited contact or spamming was used. Participants were free to decline or withdraw at any stage, and recruitment avoided any form of coercion. All organizations were treated equally in data collection and analysis. At no point were sensitive keys, cryptographic material, or detailed system configurations collected.  

\emph{Content safeguards:} All transcripts were anonymized; neither participant names nor organizational identifiers appear in published results. Findings are reported in aggregate at the thematic level (e.g., categories of usability issues) rather than specific configurations or vulnerabilities.    

\emph{Privacy \& data minimization:} Only de-identified transcripts were retained; analysis outputs emphasize themes across participants, not individual accounts. No identifiable deployment details are included in the publication.  

\emph{Operational safety:} We deliberately excluded or generalized any descriptions of signing deployments that could aid adversaries, including specific weaknesses in Sigstore’s components. Where risks were mentioned (e.g., transparency log privacy concerns, private instance deployment issues), results are presented at the level of usability factors rather than detailed attack vectors.

\vspace{0.7em}
\noindent
\textbf{Participant Welfare.}  
As mentioned earlier, all participants gave informed consent.
Potential distress was minimal, though participants might experience discomfort when reflecting on organizational challenges.
To mitigate this, questions emphasized practices and perceptions rather than failures, and responses were anonymized.
No deception was involved.

\vspace{0.7em}
\noindent
\textbf{Oversight and Consent Context.}  
The academic institution’s IRB reviewed and approved the protocol.
Prior to consenting, all participants were informed of the study's aims and the confidentiality safeguards in place.

\vspace{0.7em}
\noindent
\textbf{Decision to Proceed and Publish.}  
We judged that the expected benefits—independent evidence on identity-based signing tool usability, insights into adoption barriers, and guidance for improving software supply chain security—outweigh residual risks under the mitigations described above. To minimize adversarial risk, we publish only aggregated themes, not sensitive technical details, to ensure that our findings strengthen rather than weaken trust in signing infrastructures.

% \cleardoublepage

% \section*{Open Science}
% \textbf{Within up to one page, this appendix must list all artifacts necessary to evaluate the contribution of the paper and make clear how the review committees can access each artifact. This appendix must have exactly this title, otherwise you will risk desk rejection. }
% % optional clearing of the page
% \cleardoublepage
\clearpage
\section*{Open Science} \label{sec:OpenScience}

We acknowledge that USENIX Security has an open science policy: that authors are expected to openly share their research artifacts by default.
The research artifacts associated with this study are:
\begin{itemize}[leftmargin=12pt, rightmargin=5pt]
\ifARXIV
    \item Raw transcripts of interviews
    \fi
    \item Raw transcripts of interviews
    \item Anonymized transcripts of interviews
    \item Interview protocol
    \item Codebook
\end{itemize}

\myparagraph{Things we have shared}
The \textit{interview protocol} is a crucial part of any interview study, since it allows for the critical review of a study design as well as its replication
\ifARXIV Our full interview protocol is included in~\cref{sec:appendix-InterviewProtocol}\fi.
Since it is semi-structured, we include all of the questions asked of all subjects, as well as examples of the follow-up questions we asked.
We also share the \textit{codebook}, with codes, definitions, and example quotes that we coded for each code.
% --- see~\cref{sec:appendix-Codebooks}.

\myparagraph{Things we cannot share}
For subject privacy reasons, and for IRB compliance, we cannot share the raw transcripts.
We also choose not to share the anonymized transcripts of the interviews.
Given the high organizational ranks of many of our subjects, and the small size of the subject pool resulting from our recruiting strategy, we believe there is a high risk of de-anonymization even of anonymized transcripts.
Therefore, we do not share the anonymized transcripts.

\myparagraph{Artifact Repository} An artifact containing our interview protocol, codebook, etc. is available at: \url{https://doi.org/10.5281/zenodo.17969423}

\raggedbottom
% \cleardoublepage

%  % \pagebreak
% \section*{Ethical Considerations}
% \textbf{Within up to one page, explain the ethical considerations of your work. This appendix must have exactly this title, otherwise you will risk desk rejection. Carefully study the Ethics Guidelines before submitting your paper.}

% \section*{Open Science}
% \textbf{Within up to one page, this appendix must list all artifacts necessary to evaluate the contribution of the paper and make clear how the review committees can access each artifact. This appendix must have exactly this title, otherwise you will risk desk rejection. }
% optional clearing of the page
% \cleardoublepage

%\newpage
 % \clearpage

%-------------------------------------------------------------------------------

% \bibliographystyle{IEEEtran}
% \bibliography{bibliography/new, bibliography/references}
\bibliographystyle{plain}
\bibliography{bibliography/current}

\ifARXIV
\section*{Outline of Appendices}

\noindent
The appendix contains the following material:

\begin{itemize}[leftmargin=12pt, rightmargin=5pt]

\item \cref{sec:appendix-signing}: Traditional Software Signing Workflow.
\item \cref{sec:appendix-InterviewProtocol}: The interview protocol.
\item \cref{sec:appendix-Codebooks}: The codebooks used in our analysis, with illustrative quotes mapped to each code.

\item \cref{sec:appendix-CodeSaturation}: Code Saturation Analysis.

\item \cref{sec:appendix-OtherResults}: Other Results Omitted from the Paper Due to Space Concerns .

\item \cref{sec:appendix-SigstoreComponentsMostlyUsed}: Sigstore Components Mostly Used by our Sample Population.
\item \cref{sec:appendix-OtherUsabilityContext}: Other Effects of Organization and Signing
Tools on Identified Usability Concerns Continued from \cref{sec:org_context}
\item \cref{sec:appendix-ExpandedDemographic}: Expanded Demographic Table.
\item \cref{sec:appendix-FatorsInfluencingSigstoreAdoption}: Summary of Factors Influencing Sigstore Adoption for Sigstore Users (Prior to Adoption). 
\item \cref{sec:appendix-FatorsInfluencingSigstoreConsideration}: Summary of Factors Influencing Consideration of Sigstore Amongst Non-Sigstore Users. 
\item \cref{sec: appendix-AdditionalDiscussions}: Additional Discussions \& Implications of Results.
\item \cref{sec:appendix-SurveyAttempt}: Survey results, omitted from the main paper due to low-quality data.

\end{itemize}

\section{Traditional Signing Tools Workflow} 
\label{sec:appendix-signing}
The signature creation process begins when the maintainer’s artifact is ready for submission. The signer generates a key pair that provides them with a private key for signing their software artifact and a public key for a verifier to validate the artifact’s signature. 
Once the artifact is signed, it must be submitted to a package registry (such as PyPi, npm, or Maven) or the intended party. During this phase, the signer is responsible for making their public key available and discoverable so future users can verify the signer’s identity. In the verification phase, the verifier retrieves the signer’s public key, uses it to decrypt the signature file, and checks for equivalency with the software artifact’s hashed digest. 
We show a typical legacy key-managed package signing workflow in~\cref{fig: Signing-figT} compared to a identity-based workflow in \cref{fig: nextgen-fig}.
\begin{figure}
    \centering
    \includegraphics[width=0.99\linewidth]{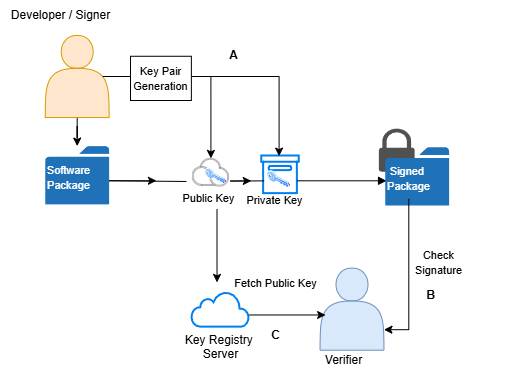}
    \caption{
    Typical workflow for software signing and verifying signatures.
    The component author packages (A) and signs (B) their software.
    The signed package (C) and public key (D) are published.
    To use a package, a user downloads it (E) and its public key (F) and verifies the signature (G).
    }
    \label{fig: Signing-figT}
    %\vspace{-70pt}
\end{figure}

\begin{figure}
    \centering
    \includegraphics[width=0.99\linewidth]{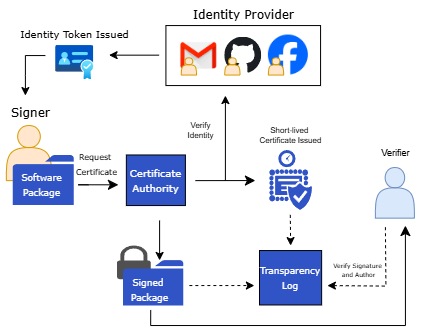}
    \caption{
    Typical workflow for identity-based software signing. The component author requests a certificate from a certificate authority, which confirms the signer's identity through an identity provider. The signature and signer's identity are recorded in a transparency log, which the verifier can monitor to confirm the validity of the signature upon downloading the signed package.
    }
    \label{fig: nextgen-fig}
    %\vspace{-60pt}
\end{figure}

\fi

\ifARXIV

\section{Interview Protocol} \label{sec:appendix-InterviewProtocol}

Here we present the full protocol, see \cref{table:InterviewProtocolCondensed}.
We indicate the structured questions (asked of all users) and some examples of follow-up questions posed in the semi-structured style. Given the nature of a semi-structured interview, the questions may not be asked exactly as written or in the same sequence, but may be adjusted depending on the flow of the conversation.

Section B provides context for interpreting our results across organizational cases. Section C lists the main questions for this study, along with any follow-up questions.

\begin{table*}
\centering
\caption{Interview Protocol. Sections A–C reflect demographics, signing use/motivation context, and signing tool usability. We analyze part C in this study and use the other themes as context.}
\label{table:InterviewProtocolCondensed}
\small
% Two columns: narrower Theme (2.8cm), wider questions (14.2cm)
\begin{tabular}{@{} >{\raggedright\arraybackslash}p{2.2cm} >{\raggedright\arraybackslash}p{15.8cm} @{}}
\toprule
\textbf{Theme} & \textbf{Description / Questions}\\
\midrule

% ============= A: Demographics (5 rows) =============
\multirow[c]{-1.5}{*}{%
  \centering
  \rotatebox[origin=c]{90}{%
    \parbox{2.8cm}{\centering \textbf{A:}\\[2pt]\textbf{Demographics}}
  }%
}
& We ask basic demographic questions to establish participant experience and context (role, years of experience, team size, and major products).\\[6pt]
& \textbf{A-1:} \textit{What best describes your role in your team? (Security engineer, Infrastructure, Software engineer, etc.)}\\
& \textbf{A-2:} \textit{What is your seniority level? How many years of experience?}\\
& \textbf{A-3:} \textit{What is the team size?}\\
& \textbf{A-4:} \textit{What are the team’s major software products/artifacts?}\\
\midrule

% ===== B: Signing Use & Context (14 rows) =====
\multirow{14}{*}{%
  \centering
  \rotatebox[origin=c]{90}{%
    % <-- only two lines for the B label -->
    \parbox{2.8cm}{\centering \textbf{B:}\\[2pt]\textbf{Signing Use \&}\\\textbf{Context}}
  }%
}
& We ask about the organizational context of signing: why and how signing is used, what motivates adoption, what infrastructure supports it, and how it fits into broader supply chain security practices. Focus is on general practices, not tooling specifics.\\[6pt]
& B-1: Why did the team choose to implement software signing?\\
& B-2: Is software signing the team’s major strategy to secure its supply chain? Any complementary efforts?\\
& B-3: Are these strategies influenced by regulations or standards?\\
& B-4: What challenges did your team face while integrating signing into supply chain practices?\\
& B-5: How do team members contribute to the implementation of signing? Roles/responsibilities?\\
& B-6: Do you believe signing (if implemented correctly) is sufficient to secure a supply chain?\\
& B-7: How does the presence of a signature influence dependency selection?\\
& B-8: How is authenticity of a signature verified in your process?\\
& B-9: What influences your team to discontinue a dependency? Does signature status play a role?\\
& B-10: How does your team manage third-party dependency security and vulnerabilities?\\
& B-11: Is signing a requirement for developers and contributors? If so, in which parts of the workflow?\\
& B-12: How is signing applied in your build process?\\
& B-13: How is signing applied to artifacts and deployment (binaries, repositories, pipelines)?\\
\midrule

% ===== C: Signing Tool Usability (7 rows) =====
\multirow[c]{8}{*}{%
  \centering
  \rotatebox[origin=c]{90}{%
    % centered C label
    \parbox{2.8cm}{\centering \textbf{C:}\\[2pt]\textbf{Signing Tool}\\\textbf{Usability}}
  }%
}
& We ask about the signing tools themselves: what tools are in use, factors influencing adoption, challenges, and considerations of alternatives. Each question includes a brief explanation.\\[6pt]
& \textbf{C-1:} \textit{What software signing tool does the team use?} (Understanding current tool choices and components in practice.)\\
& \textbf{C-2:} \textit{What factors did the team consider before adopting this tool over others?} (Identifying decision-making drivers.)\\
& \textbf{C-3:} \textit{What was the team’s previous signing practice before the current tool?} (Capturing historical evolution of practices.)\\
& \textbf{C-4:} \textit{How does the team implement this tool (which components are in use)?} (Understanding deployment in workflows.)\\
& \textbf{C-5:} \textit{What challenges or limitations has the team faced with this tool? How did you cope with these?} (Surfacing usability pain points and coping strategies.)\\
& \textbf{C-6:} \textit{Has the team considered switching tools? If yes, which alternatives?} (Exploring sustainability of current tool choices.)\\
\bottomrule
\end{tabular}
\end{table*}

\ifARXIV
\section{Codebooks} \label{sec:appendix-Codebooks}

\cref{tab:codebook} describes our codebook, with code definition, and example quote tagged with that code.

{
	%\arraystretch{1.5}
	\begin{table*}[!t]
		\centering
		\caption{
   Excerpts from the final codebook we used for our analysis.
		}
		\scriptsize
        \small
		\begin{tabular}{p{2.0in}p{2.0in}p{2.5in}}\hline
            \toprule
             \textbf{Code} & \textbf{Definition}& \textbf{Sample Quote}  \\
            \midrule
             & {}  \\
            Sigstore Criticisms & Criticisms and weaknesses associated with implementing Sigstore for signing. & P3: "We do have some concerns about the way Fulcio operates as its own certificate authority. So we've been looking at things like OpenPubkey as something that removes that intermediary and would allow you to do identity-based signing directly against the OIDC provider. " \\
             & &\\
            \textbf{} & {} & {} \\
            Sigstore Improvements & Suggested improvements for Sigstore.   &  P9: "I would say probably integrations within different platforms, so say you use GitLab or GitHub or Jenkins or something like that. Try to leverage some type of integration to make that, I guess, less friction for the developers to implement, something like that." \\
             & &\\
             \textbf{} & {} & {} \\
            Sigstore Strengths&Strengths of Sigstore and reasons why subjects prefer using Sigstore. & P7: 
"The convenience and the good user experience is probably what keeps us using Sigstore " \\
            & &\\
             \textbf{}&	\\
              Sigstore Alternatives Considered &Alternatives of Sigstore that were or are being considered by the subject (for subjects whose primary signing implementation is Sigstore).& P5: "We are also looking at other new alternatives that are coming up, like OpenPubkey. Docker just released a new standard called OpenPubKey, which is trying to solve this issue where they will not require you to have infrastructures and stuff."
             \\
             & &\\
            \textbf{} & {} & {} \\
              Sigstore Components Used & Components of Sigstore (Cosign, Fulcio, etc.) being used by the subject.& P2: "Mostly fulcio is what's integrated into our product." \\
            & &\\
             % \textbf{} & {} & {} \\
             %  Sigstore Criticisms& Criticisms and weaknesses associated with implementing SigStore for signing & S3: "We do have some concerns about the way Fulcio operates as its own certificate authority. So we've been looking at things like OpenPubkey...removing that intermediary...to do identity-based signing directly against the OIDC provider."\\
             %  & &\\
            \textbf{} & {} & {} \\
               Notary Criticism& Criticisms and weaknesses Associated with Notary -- promoting users to stop using them. & "P5: "We were using Notary, and Notary only verifies that the image was signed using a specific key, but you cannot verify that who is the owner of the key." \\
              & &\\
            % \textbf{} & {} & {} \\
            %   Notary Strength & Strengths associated with Notary.& S5: "Besides that, if you use something other than Sigstore like Notary, then you might not face that issue."  \\
            %  & &\\
%              \textbf{} & {} & {} \\
%               Hashi Vaulte/Spiffe&Strengths and weaknesses associated with use of Spiffe and Hashi/Vaulte.& S2: 
% "The stuff that Hashi's doing with Vault, they do a very good job with it. And then the SPIFFE/SPIRE community, they do a very good job of it. They're based upon Google LOAS, so internal Google system. So there's a lot of different ways that we know how to distribute keys in a secure and safe way."  \\
            & &\\
%              \textbf{} & {} & {} \\
%              Other Unnamed Signing Implementation Strength& Strengths and reasons why internally implemented signing tools were used by the subject.& S11: 
% "I think any tool that's being used has to go through a security review process. So the security review has been done and we understand what this tool is, and we have enabled the teams to use this tool internally. And this is within our infrastructure, ... And then we enable teams to use this infrastructure that's approved by the central security team. So that's how we approach it. ... We want to centralize signing. We want to make sure we have good key management practices and then use NHSM or some other infrastructure to generate these keys securely..."\\
            \bottomrule
		\end{tabular}
		\label{tab:codebook}
	\end{table*}
}
\else

\fi
\fi

\section{Code Saturation} \label{sec:appendix-CodeSaturation}

\begin{figure}[hbt!]
    % \vspace{-13.0pt}
    \centering    \includegraphics[width=0.99\linewidth]{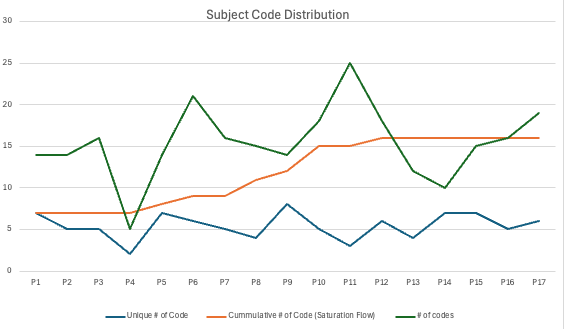}
    \caption{ 
    Saturation curve.
    Interviews are plotted in the order in which they were conducted.
    Each interview covered many topics in detail (orange line, blue line).
    However, as the green line shows, our results saturated (\ie stopped observing new perspectives) around interview 12.
    \JD{Is this the saturation curve for the full codebook (including USENIX) or just the SOUPS codebook?}
    \KC{Just soups}
     }
    \label{fig: saturation}
    % \vspace{-10.50pt}
\end{figure}

% \vspace{5.0pt}
We present a saturation analysis of our studies in ~\cref{fig: saturation}. The orange trendline (saturation curve) was measured by identifying the cumulative unique codes present in each interview. We observe saturation at the 12th interview, indicating that no new unique codes were identified after participant 12. Additionally, the green trendline (number of codes) shows that the lowest number of codes was obtained from subject P4. This is also reflected in the blue trendline (unique number of codes).

% \vspace{2em}
\FloatBarrier   % <-- forces Appendix B floats to appear before moving on

\ifARXIV

\else
\section{Technical Report} 
\label{sec:appendix-TechnicalReport}
A extended version of our paper is available at \url{https://arxiv.org/abs/2503.00271}.
The primary difference is its appendices, which contain the following information:

% \noindent\textbf{Additional contents.}
% For transparency and reuse, the appendix includes:
\begin{itemize}[leftmargin=12pt, rightmargin=5pt]
  \item Appendix A: Traditional software signing workflow (diagram and narrative description).
  \item Appendix B: The full interview protocol used in our study.
  \item Appendix C: The codebooks used in our analysis, with illustrative quotes mapped to each code.
  \item Appendix D: Code saturation analysis.
  \item Appendix E: Survey results that were omitted from the main paper due to low-quality data.
  
    \item Appendix E.1: Sigstore components mostly used by our sample population.
    \item Appendix E.2: Other results omitted from the paper due to space constraints.
    \item Appendix E.3: Expanded demographic table.
    \item Appendix E.4: Summary of factors influencing Sigstore adoption for Sigstore users (prior to adoption).
    \item Appendix E.5: Summary of factors influencing consideration of Sigstore among non-Sigstore users.
    \item Appendix E.6: Additional discussions and implications of results.
    \item Table 13: Driver labels (Push/Pull/Barrier) mapped to concrete codes with factor tags (T/P/O/M) and example participant IDs.
\end{itemize}

% \FloatBarrier   % <-- forces Appendix B floats to appear before moving on

\fi

\ifARXIV

\fi

\ifARXIV

\section{Other Results}
\label{sec:appendix-OtherResults}
In this appendix we present a continuation of the results presented in  \cref{sec:Results}.

\subsection{Sigstore Components Most Commonly Used by Our Sample Population} 
\label{sec:appendix-SigstoreComponentsMostlyUsed}

See \cref{tab: component_usage}, which highlights the Sigstore components most commonly used by our study participants.

\begin{table}[]
\scriptsize
% \small
\caption{
Sigstore Functionalities Used by Subjects.
\JD{Add another sentence giving interpretation.}
}
\begin{tabular}{lc}
\toprule
% \multicolumn{2}{l}{\textbf{Sigstore Functionalities Utilized by Subjects}} \\ \hline
\textbf{Tool}                                 & \textbf{\# of Subjects}    \\ 
\toprule
Cosign (Signing \& Verification CLI)                                        & 9                          \\
Certificate Authority (Fulcio)                & 9                         \\
Keyless Signing \& Identity Manager (OIDC)                       & 9                          \\
Transparency Log (Rekor)                      & 8                          \\
Gitsign (Commit Signing)                                       & 4                          \\
Component Customization \& Local Sigstore Deployment     & 2                          \\
% TSA                                           & 1  
\bottomrule
\end{tabular}%
\label{tab: component_usage}
\end{table}

\subsection{Other Effects of Organization and Signing Tools on Identified Usability Concerns}
\label{sec:appendix-OtherUsabilityContext}

In this appendix, we present a continuation of the results shown in \cref{fig:heatmapsRQ1}, highlighting the organizational and usage contexts reported by participants in medium-scale organizations.

\begin{figure}
    \centering
    \includegraphics[width=0.99\linewidth]{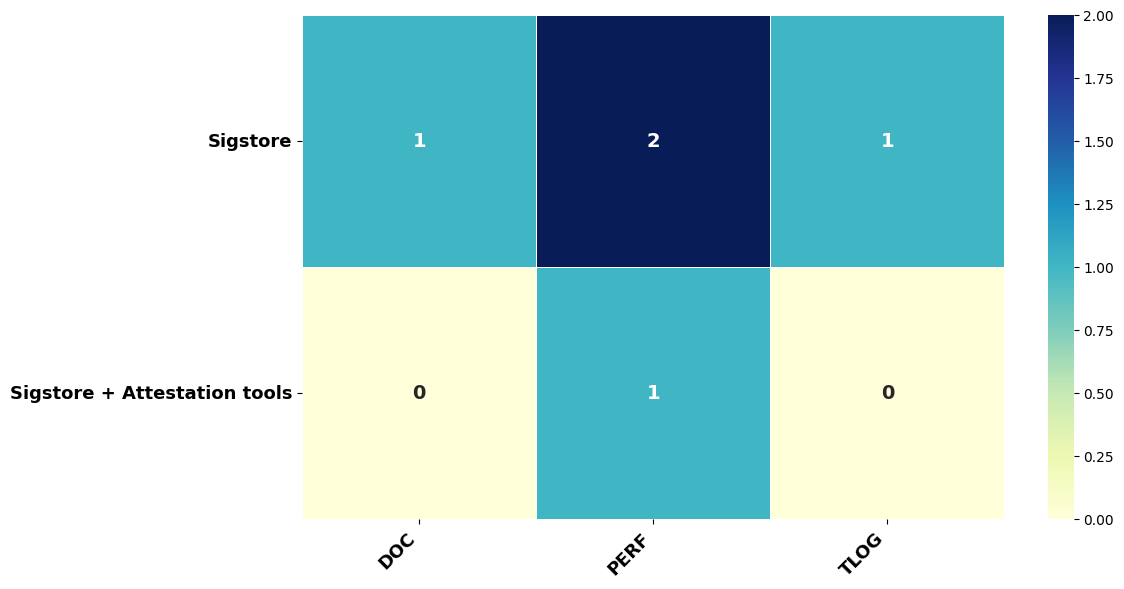}
    \caption{
    Heatmap showing the organizational usability context for subjects from medium scale organizations (4 subjects). Similar to the one showed for small and large-scale organizations in~\cref{fig:heatmapsRQ1}.
    In comparing to~\cref{fig:heatmapsRQ1}, note that some columns are omitted in this figure because no subjects mentioned those aspects.
    }
    \label{fig: appendix-HeatmapMediumfig}
    %\vspace{-60pt}
\end{figure}

\subsection{Expanded Demographics Table}
\label{sec:appendix-ExpandedDemographic}

See \cref{tab:AppendixSubjects} for the extended participant demographics table, which includes interview duration (in minutes).
%table 2:
{
	%\arraystretch{1.5}
	\begin{table}[!t]
	\centering
    \caption{
    Subject Demographics.
    For anonymity, we used generic job roles, not specific titles.
    ``\textit{Senior management}'' refers to senior managers, directors, and executives;
    ``\textit{Technical leader}'' to senior, lead, partner, and principal engineers;
    and
    ``\textit{Engineer}'' and ``\textit{Manager}'' to more junior staff.
    We highlight the type of software subject's immediate team produces. \textit{*} refers to cases where subject's team built other types of software in addition to the main stated. \textit{O} refers to cases where subject belonged to multiple teams with other products than the one stated.
    }
    
    \scriptsize
	\begin{tabular}{llclc}
        \toprule
       \textbf{ID} & \textbf{Role} & \textbf{Experience} & \textbf{Software Type} & \textbf{Duration (min)} \\
        \midrule
        P1  & Research leader    & 5 years   & Internal POC software                     & 55 \\
        P2  & Senior mgmt.       & 15 years  & SAAS security tool*\textsuperscript{O}    & 45 \\
        P3  & Senior mgmt.       & 13 years  & SAAS security tool*\textsuperscript{O}    & 43 \\
        P4  & Technical leader   & 20 years  & Open-source tooling                       & 33 \\
        P5  & Engineer           & 2 years   & Internal security tooling                 & 46 \\
        P6  & Technical leader   & 27 years  & Internal security tooling*\textsuperscript{O} & 68 \\
        P7  & Manager            & 6 years   & Security tooling*\textsuperscript{O}      & 48 \\
        P8  & Technical leader   & 8 years   & Internal security tooling*                & 57 \\
        P9  & Engineer           & 2.5 years & SAAS security*                            & 55 \\
        P10 & Engineer           & 13 years  & SAAS security*                            & 45 \\
        P11 & Technical leader   & 16 years  & Firmware*                                 & 57 \\
        P12 & Technical leader   & 4 years   & Consultancy                               & 65 \\
        S13 & Senior mgmt.       & 16 years  & Internal security tool                    & 58 \\
        P14 & Research leader    & 13 years  & POC security Software*                    & 52 \\
        P15 & Senior mgmt.       & 15 years  & Internal security tooling                 & 42 \\
        P16 & Senior mgmt.       & 15 years  & SAAS                                      & 60 \\
        P17 & Manager            & 11 years  & Security tooling                          & 52 \\
        \bottomrule
	\end{tabular}
	\label{tab:AppendixSubjects}
\end{table}

}

\fi

\ifARXIV
\subsubsection{Organizational Demographics} \label{sec:appendix-Demographics}
We summarize additional organizational demographic information of our subjects in \cref{tab:orgSummary}. This table provides further context for the subject's organizational size, type of software product, and how many of our subjects belonged to each organization.

{
\renewcommand{\arraystretch}{1.3}
\begin{table}
    \centering
    \caption{
    Organizational Demographics. To enhance anonymity, we only highlight the number of subjects in each organizational category. Letters A-L are used to depict each organization.
    % See~\cref{sec:appendix-InterviewProtocol} for full protocol.
    % Topic D is not analyzed in this paper.
   }
   \small
    \scriptsize
    \begin{tabular}{
    % p{0.13\linewidth}p{0.5\linewidth}
    p{0.3\linewidth}p{0.6\linewidth}
    }
    \toprule
        \textbf{Type}& \textbf{Breakdown (\#Organizations|\#Subjects)} \\
        \toprule
         Organizational Size (Employee Size)& {
            Small (<100) (4/6), Medium (<1500) (3/4), Large (>1500) (6/7)
         } \\
        % \midrule
         Product Area & {
             Digital technology (1/3), SSC Security (2/4), Social technology (1/1), Dev tools (1/1), Telecommunications (1/1), Cloud security (2/3), Aerospace Security (1/1), Internet services (2/2), Cloud + OSS Security (1/1), Cloud + Dev toools (1/1)
         } \\
         \midrule
        Subject Distribution& {
             A (2), B (3), C (1), D (1), E (2), F (1), G (1), H (1), I (1), J (1), J (1), K (1), L (1), M(1)
         }  \\
        \bottomrule
    \end{tabular}
   \label{tab:orgSummary}
\end{table}
}
\else

\fi

\ifARXIV
\subsection{Factors Influencing Sigstore Adoption (Sigstore Users)} \label{sec:appendix-FatorsInfluencingSigstoreAdoption}

In this appendix, we present a tabular organization of the reasons given by Sigstore users for their teams’ and organizations’ adoption of Sigstore, structured according to Cresswell \etal’s four-factor framework.

{
	%\arraystretch{1.5}
	\begin{table}[!t]
		\centering
		\caption{
		  Reasons Practitioners Choose or Switch to Sigstore Before Adoption.
          Practitioners' decisions are driven primarily by human/social factors—such as community contribution, frustration with legacy tools, and usability issues—supplemented by \sigs technological capabilities and reinforced by macroenvironmental pressures like regulations and trust in the CNCF ecosystem.
          \JD{Move this to appendix.}
		}
		\small
            \scriptsize
		\begin{tabular}{ p{0.60\linewidth}p{0.30\linewidth}}
            \toprule
            \textbf{Topics \& Associated Examples  } & \textbf{Subjects} \\
            \midrule
            \textbf{\textit{Human/Social Factors}} &  \\
            \textbf{Practitioners Contribute to Sigstore} & {6} -- P2, P4, P6, P7, P9, P14 \\
            \textbf{GPG Issues} & {4} \\
             Low adoption rates &  P1, P14\\
             Steep learning curve & P12, P16\\
            \textbf{Notary Issues} & {1} \\
              Non-demand from customers & P12 \\
             % Signing Workflow \& Verfication & --- \\
             % Setting up with automated CI/CD actions & -- \\
             % No key distribution problems & --\\
             \midrule
            \textbf{\textit{Technological Factors}} &  \\
            \textbf{Available Sigstore Functionalities} & {3} -- P1, P5, P14 \\
             A transparency log, etc & P5, P14 \\
            Compatibility to other Tools & P1 \\
             % No key distribution problems & --\\
            \textbf{GPG Issues} & {5} \\
             Key management issues \& Other usability concerns & P1, P6, P12, P16, P17\\
             Compatibility with newer technology & P16\\
            \textbf{Notary Issues} & {2} \\
              Compatibility with other tools & P12 \\
              Lack of regular updates & P12 \\
              Key \& Identity Management & P5 \\
            \textbf{Proprietary Tool Issues} & {1} \\
            Difficult to setup & P15 \\
             \midrule
            \textbf{\textit{Macroenvironmental Factors}} &  \\
            \textbf{Regulation \& Standards (Cross-factor)}- \textit{M \& O} & {4} -- P5, P6, P15, P17 \\
             % vv & -- \\
             % \sigs Compliance with Standards & --- \\
             % No key distribution problems & --\\
            \textbf{Large Sigstore User Community} & {1} -- P9 \\
            \textbf{Inherent Trust of Creators} & {1} \\
            Trust of CNCF products & P3 \\
            \bottomrule
		\end{tabular}
		\label{tab:rq2}
	\end{table}
}

\subsection{Factors Influencing Sigstore Consideration (Non-Sigstore Users).}
\label{sec:appendix-FatorsInfluencingSigstoreConsideration}

Here, we summarize the factors that non-Sigstore users say influence them and their organizations to consider adopting Sigstore. See \cref{tab:non_sigstore_reasons_for}.

\input{misc/data/non-sigstore-users-1}

% \subsection{Additional Discussions \& Implications of Results}
% In addition to the implications of our research presented in our discussion section \cref{sec:Discussion}, we present additional implications for our results of our work in the following points

% \subsubsection{Generalizability of Findings to Other Classes of Tools in Software Engineering }

% Our findings corroborated number of factors that has been presented in the literature about adoption factors especially in general open source tools, but our result also found some factors not yet covered that could generalize to this class of tools.
% Example, compatibility with existing technology, ease of use of user interface, etc. are factors that have been hinted in earlier work (with little attention to signing next gen tools)

% Our results also show a possibility of a migration pathway model that may affect the adoption of open source tool. This is exactly true as some non-sigstore users admitted considerations and even use of smaller identity-based tool integrations to solve tooling issues eg openpubkey and TUF for key delivery and management. This indicates an ... 

\subsection{Additional Discussions \& Implications of Results}
\label{sec: appendix-AdditionalDiscussions}

In addition to the implications discussed in \cref{sec:Discussion}, we outline further implications of our findings below.
\iffalse
\subsubsection{Generalizability of Findings to Other Classes of Tools in Software Engineering}

Our findings reinforce several adoption factors noted in prior literature on open source tools—such as ease of use, compatibility with existing technologies, and integration flexibility—which were especially salient for identity-based signing tools like Sigstore. They also highlight additional considerations that may generalize to other classes of secure software engineering tools example the role played by macroenvironmental factors like the tooling community, and regulations.

Furthermore, our results suggest adoption patterns not widely discussed in existing research. In particular, we observe the emergence of a migration pathway model, where teams incrementally adopt identity-based tools. Several non-Sigstore users reported experimenting with or partially integrating smaller solutions—such as OpenPubKey and TUF—to meet specific needs like key delivery and management. This suggests a stepwise transition process in tooling adoption that may be common across emerging tool ecosystems. This model can also be important to understand and further propose a streamlined organizational adoption pathway for software engineering tools.
\fi

\subsubsection{The Role of Community in Software Signing Tool Adoption}
Many participants cited their involvement in the Sigstore community as a key reason for adopting the tool (\cref{tab:rq2}). While this may partly reflect our sampling, half of the non-Sigstore users who adopted external tools also emphasized community affiliation as a motivator.
% This illustrates the influence of macroenvironmental factors—particularly community engagement—as captured in our usability framework. 
Challenges like setting up private Sigstore instances or accessing usage knowledge were frequently linked to the absence of sufficient community support (\cref{sec: sigstore_weakness}). 
Other community-related influences, such as the size of the user community and trust in the tool creators, also shaped adoption decisions.

These findings echo prior work on the importance of community in developer tool adoption \cite{sharp819967}. Toolmakers should actively foster supportive, trustworthy communities around their tools, as these networks can be pivotal to both adoption and long-term sustainability.
\fi

\ifARXIV

\subsubsection{Push-Pull-Barrier Definition}

In this appendix, we map our generated themes and subthemes to the Push, Pull, and Barrier factors from the TRL framework by Bansal \etal~\cite{bansal2005migrating}, as described in the results section. The table defines the driver type, corresponding theme, and examples.
%We define the driver type, theme from each of the (tab:).
As this analysis was highly objective, a single rater performed this mapping.
The table is here:~\cref{tab:rq2_codes_drivers}.

\begin{table*}[!t]
\centering
\scriptsize
\small
\caption{Driver labels (Push/Pull/Barrier) mapped to concrete codes with factor tags (T/P/O/M) and example participant IDs.}
\resizebox{\textwidth}{!}{%
\begin{tabular}{p{0.09\linewidth}p{0.15\linewidth}p{0.46\linewidth}p{0.06\linewidth}p{0.12\linewidth}}
\toprule
\textbf{Driver} & \textbf{Theme} & \textbf{Code} & \textbf{Factor} & \textbf{Examples}\\
\midrule
Barrier & Policy/ownership & Centralize org code signing internally & O & P8\\
Barrier & Policy/ownership & Organization’s open-source policy constraints & O & P8\\
Barrier & Policy/ownership & Organizational use cases don’t fit Sigstore & O & P8\\
Barrier & Policy/ownership & Tool owned by organization & O & P10\\
Barrier & Risk/privacy/trust & Low trust in Sigstore maintainers & M & P13\\
Barrier & Risk/privacy/trust & Sigstore privacy concerns & T & P13\\
Barrier & Risk/privacy/trust & Third-party security risk concerns & T & P11, P8\\
Pull & Community/engagement & Practitioners contribute to Sigstore & H & P2, P4, P6, P7, P9, P14\\
Pull & Macro pressures & Large Sigstore user community & M/H & P9\\
Pull & Macro pressures & Regulation \& standards alignment & M/O & P5, P6, P15, P17\\
Pull & Macro pressures & Trust in CNCF/creators & M & P3\\
Pull & Sigstore capabilities & Compatibility with other tools/CI & T & P1\\
Pull & Sigstore capabilities & Transparency log \& related features & T & P5, P14\\
Push & General & Key \& identity management pain & T & P5\\
Push & GPG issues (human) & Low adoption rates & H & P1, P14\\
Push & GPG issues (human) & Steep learning curve & H & P12, P16\\
Push & GPG issues (tech) & Compatibility with newer technology & T & P16\\
Push & GPG issues (tech) & Key management \& other usability pain & T & P1, P6, P12, P16, P17\\
Push & Notary v1 issues & Engineers don’t understand tool & P & P11\\
Push & Notary v1 issues & Key management & T & P10\\
Push & Notary v1 issues & Lack of in-built signing policy & T & P10\\
Push & Notary v1 issues & Lack of regular updates & T & P12\\
Push & Notary v1 issues & Compatibility problems & T & P12\\
Push & Proprietary/Internal tool issues & Difficult to set up & T & P15\\
Push & Proprietary/Internal tool issues & Engineers don’t understand current tool & P & P11\\
Push & Proprietary/Internal tool issues & Integration with other tools/platforms & T & P8, P11\\
Push & Proprietary/Internal tool issues & Non-transparent signing infrastructure & T & P11\\
Push & Proprietary/Internal tool issues & Non-unified implementations across teams & O & P8, P11\\
Push & Proprietary/Internal tool issues & Security concerns about signing infrastructure & T & P11\\
Push & Proprietary/Internal tool issues & Signer ID management problems & T & P11\\
Push & Proprietary/Internal tool issues & Unclear usage info/docs & T & P11\\
\bottomrule
\end{tabular}%
}
\label{tab:rq2_codes_drivers}
\end{table*}

\fi

\ifARXIV
\subsection{Survey Attempt} \label{sec:appendix-SurveyAttempt}
\SO{kelechi you can fill in}
% In addition to our interview, we developed a survey that was distributed to Sigstore users (distributed At the 2024 Sigstorecon conference and other post-conference communications to the Sigstore user community) to collect perspectives from those who might not have had the time for an interview. 
In addition to our interviews, we developed a survey distributed to Sigstore users at the 2024 SigstoreCon conference and through post-conference communications with the Sigstore user community. This aimed to gather perspectives from those who might not have had time for an interview.
We exclude this data from our work due to several factors that compromise the quality of the results; these factors include respondents who completed the survey in an unrealistically short amount of time and those who failed to complete the entire survey.

Here, we summarize the answers to an excerpt of the questions responded to in our survey:

\textbf{What signing tool did your team use prior to the introduction of Sigstore?} Among 7 participants who utilize Sigstore, 3/7 responded that they use PGP/GPG-based signing and 4/7 did not use any toolings prior to using Sigstore.

\textbf{Which of the following factors have influenced your team’s use of signing? (select all):} Among 7 participants who utilize Sigstore, 4/7 selected "Organizational policies," 4/7 with "Best Practices/Other standards and recommendations" and 3/7 selected "Awareness of prominent SSC attack."

\textbf{Which Sigstore components does your team use? (select all):} Among 7 participants who utilize Sigstore, 3 responded to this question, and all 3 participants selected Cosign, Fulcio, and Rekor.

\textbf{In which of the following ways have you implemented Sigstore?} Among 7 participants who utilize Sigstore, 3 responded to this question, with 2 responding that they use a private deployment of Sigstore and 1 responding that they use a public instance of Sigstore.

\textbf{Do you use or require Sigstore (signing) at any of the following stages of your product?} Among 7 participants who utilize Sigstore, 3 responded to this question. All 3 participants shared the "Build/Deployment/final product" stage as part of their answer.

\textbf{How do you verify Sigstore signatures? (select all):} Among 7 participants who utilize Sigstore, 3 responded to this question. All 3 participants shared selecting certificate Verification as part of their answer. 2/3 selected that they check Rekor logs.
\fi

%-------------------------------------------------------------------------------

%%%%%%%%%%%%%%%%%%%%%%%%%%%%%%%%%%%%%%%%%%%%%%%%%%%%%%%%%%%%%%%%%%%%%%%%%%%%%%%%
\end{document}

%% file: misc/typesetting.tex
\usepackage{amsmath,amsfonts}
\usepackage{algorithmic}
\usepackage{graphicx}
\usepackage{textcomp}
\usepackage{xcolor}
\usepackage{booktabs} % For formal tables \toprule
\usepackage{xspace} % Put an \xspace within numeric macros
\usepackage[normalem]{ulem}
\usepackage{makecell}
\usepackage{tcolorbox}
\usepackage{enumitem}
\usepackage{siunitx}
\usepackage[vskip=1em,font=itshape,leftmargin=2em,rightmargin=2em]{quoting}
\usepackage{svg}
\usepackage{listings}
\usepackage{subfigure}
\usepackage{xcolor}
\PassOptionsToPackage{hyphens}{url}\usepackage{hyperref}% Defining where it is okay to wrap a url to a new line. For bibliography.

\usepackage{soul}
\usepackage{hyperref}
\usepackage{cleveref}

\ifDEBUG
   \newcommand{\JD}[1]{\textcolor{purple}{[Jamie says: #1]}}
    \newcommand{\KC}[1]{\textcolor{blue}{[Kelechi says: #1]}}
    \newcommand{\SC}[1]{\textcolor{olive}{[Sophie says: #1]}}
    
    \newcommand{\ST}[1]{\textcolor{orange}{[Santiago says: #1]}}
    \newcommand{\SO}[1]{\textcolor{pink}{[Sofia says: #1]}}
    
    \newcommand{\TODO}[1]{\hl{#1}}
\else
    \newcommand{\JD}[1]{}
    \newcommand{\KC}[1]{}
    \newcommand{\SC}[1]{}
    \newcommand{\AK}[1]{}
    \newcommand{\ST}[1]{}
    \newcommand{\SO}[1]{}
    \newcommand{\TODO}[1]{#1}
\fi

\newcommand{\myparagraph}[1]{\vspace{0.08cm}\noindent\hspace{0.1cm} \ul{#1:}}

\usepackage{balance} % Put \balance on the last page
\usepackage{amsmath}
\usepackage{cleveref}

\crefformat{section}{\S#2#1#3}
\crefname{figure}{Figure}{Figures}
\crefname{appendix}{Appendix}{Appendices}
\crefname{table}{Table}{Tables}
\crefname{table2}{Table}{Tables}
\crefname{algorithm}{Algorithm}{Algorithms}
\crefname{listing}{Listing}{Listings}
\crefname{theorem}{Theorem}{Theorems}
\crefname{thm}{Theorem}{Theorems}
\crefname{lemma}{Lemma}{Lemmata}
\crefname{equation}{Eqt.}{Eqts.}
\crefformat{Grammar}{Grammar #1}

\usepackage{enumitem}
\usepackage{booktabs}

\usepackage{makecell}
\usepackage{multirow}
\usepackage[T1]{fontenc}

\newcommand{\ie}{\textit{i.e.,} }
\newcommand{\eg}{\textit{e.g.,} }
\newcommand{\etal}{\textit{et al.}\xspace}
\newcommand{\etals}{\textit{et al.'s}\xspace}

\usepackage{enumitem}

\newcommand{\myinlinequote}[1]{\emph{``#1''}}

\usepackage{quoting}

\quotingsetup{
  leftmargin=0.15cm,
  rightmargin=0.15cm,
  font=italic,
  vskip=0.10cm,        % minimal vertical space before/after
  indentfirst=false
}

\newcommand{\ParticipantQuote}[2]{
    \begin{quoting}
    \emph{\textbf{#1}: ``#2''} 
    \end{quoting}}

\DeclareUnicodeCharacter{2060}{\nolinebreak}

\newcounter{finding}

%% file: data/data.tex
% ============= MAVEN CENTRAL =================
\newcommand{\maven}{Maven Central\xspace}
\newcommand{\mavenPackages}{{\TODO{30}}\xspace}
\newcommand{\mavenOrganizations}{{\TODO{16}}\xspace}

\newcommand{\mavenReevaluateSelected }{{\TODO{10\%}}\xspace}

\newcommand{\ssc}{Software Supply Chain\xspace}
\newcommand{\sscsm}{software supply chain\xspace}
\newcommand{\sscs}{software supply chains\xspace}
\newcommand{\oss}{Open Source Software\xspace}
\newcommand{\interviews}{18\xspace}
\newcommand{\subjects}{18\xspace}
\newcommand{\sign}{Software signing\xspace}
\newcommand{\signs}{software signing\xspace}
\newcommand{\signature}{Software Signature\xspace}
\newcommand{\orgs}{13\xspace}
\newcommand{\seps}{software engineering processes\xspace}
\newcommand{\sep}{Software Engineering Process\xspace}
\newcommand{\sscfm}{software supply chain factory model\xspace}
\newcommand{\extContribution}{10\xspace}
\newcommand{\reqExternal}{5\xspace}
\newcommand{\intSign}{14\xspace}
\newcommand{\sig}{Sigstore\xspace}
\newcommand{\sigs}{Sigstore's\xspace}
\newcommand{\toolings}{\textit{tooling}\xspace}
\newcommand{\toolingl}{\textit{Tooling}\xspace}
\newcommand{\themes}{\xspace}
\newcommand{\subthemes}{\xspace}

\newcommand{\Subjectone}{\emph{P1 \small{\textit{(POC software, digital technology)}}}\xspace}
\newcommand{\Subjecttwo}{\emph{P2 \small{\textit{(SAAS, SSC security tool)}}}\xspace}
\newcommand{\Subjectthree}{\emph{P3 \small{\textit{(SAAS, SSC security tool)}}}\xspace}
\newcommand{\Subjectfour}{\emph{P4 \small{\textit{(open source tooling, cloud dev tool)}}}\xspace}
\newcommand{\Subjectfive}{\emph{P5 \small{\textit{(internal security tool, cloud security)}}}\xspace}
\newcommand{\Subjectsix}{\emph{P6 \small{\textit{(security tooling, digital technology)}}}\xspace}
\newcommand{\Subjectseven}{\emph{P7 \small{\textit{(security tools, cloud security)}}}\xspace}
\newcommand{\Subjecteight}{\emph{P8 \small{\textit{(internal security tool \& cloud APIs, internet services)}}}\xspace}
\newcommand{\Subjectnine}{\emph{P9 \small{\textit{(SAAS, SSC security)}}}\xspace}
\newcommand{\Subjectten}{\emph{P10 \small{\textit{(SAAS, dev tools)}}}\xspace}
\newcommand{\Subjecteleven}{\emph{P11 \small{\textit{(firmware \& testing software, social technology)}}}\xspace}
\newcommand{\Subjecttwelve}{\emph{P12 \small{\textit{(consultancy, cloud OSS security)}}}\xspace}
 \newcommand{\Subjectthirteen}{\emph{P13 \small{\textit{(internal security tool, telecommunication)}}}\xspace}
\newcommand{\Subjectfourteen}{\emph{P14 \small{\textit{(POC software \& security tool, cloud security)}}}\xspace}
\newcommand{\Subjectfifteen}{\emph{P15 \small{\textit{(internal security tooling, aerospace security)}}}\xspace}
    \newcommand{\Subjectsixteen}{\emph{P16 \small{\textit{(SAAS, SSC security tool)}}}\xspace}
\newcommand{\Subjectseventeen}{\emph{P17 \small{\textit{(security tooling, internet service)}}}\xspace}

%% file: misc/data/table-nonsig-2.tex
\begin{table}[ht]
\centering
\caption{Non-Sigstore Users: Reasons for Not Adopting Sigstore. This table summarizes the reasons non-users gave for not adopting Sigstore, grouped by their current signing tool and mapped to our usability framework theme.}
\vspace{5pt}
\scriptsize
\label{tab:non_sigstore_reasons_against}
\resizebox{\columnwidth}{!}{%
\begin{tabular}{p{0.15\linewidth} p{0.12\linewidth} p{0.60\linewidth}}
\toprule
\textbf{Tool} & \textbf{Subjects} & \textbf{Reason (Topic)} \\
\midrule
Internal tool & P8, P11 & (1) Third‐party security risk from tool \textbf{T} -- P11, P8 \\
{} & {} & (2) Organizational use cases \textbf{O} -- P8 \\
{} & {} & (3) Centralizing organization's code signing \textbf{O} -- P8 \\
{} & {} & (4) Organization's open source policy \textbf{O} -- P8 \\
\midrule
Notary-V1 & P10 & (1) Tool owned by organization \textbf{O} -- P10 \\
\midrule
PGP & P13 & (1) Sigstore privacy concerns \textbf{T} -- P13 \\
{} & {} & (2) Trust in Sigstore’s maintainers \textbf{M} -- P13 \\
\bottomrule
\end{tabular}
} % end resizebox
\end{table}

\iffalse
\begin{table}[ht]
\centering
\small
\caption{Non Sigstore Users: Reasons for Not Adopting Sigstore. This table summarizes the reasons non-users gave for not adopting Sigstore, grouped by their current signing tool and mapped to our usability framework theme.}
\normalsize
\small
\scriptsize
\label{tab:non_sigstore_reasons_against}
\begin{tabular}{p{0.15\linewidth} p{0.12\linewidth}p{0.60\linewidth}}
\toprule
\textbf{Tool} & \textbf{Subjects}  & \textbf{Reason (Topic)} \\
\midrule
Internal tool      & P8, P11                         &(1) Third‐party security risk from tool \textbf{T} -- P11, P8 \\
{} & {}  &(2) Organizational Use cases \textbf{O} -- S8 \\
{} & {}  &(3) Centralizing Organization's code signing \textbf{O} -- P8 \\
{} & {}  &(4) Organization's Open Source Policy \textbf{O} -- P8 \\
\midrule
Notary-V1       & P10                             & (1) Tool owned by Organization \textbf{O} -- S10\\
\midrule
PGP           & P13                        & (1) Sigstore Privacy concerns \textbf{T} -- P13 \\
{} & {}  &(2) Trust in Sigstore’s Maintainers \textbf{M} -- P13 \\
\bottomrule
\end{tabular}
\end{table}
\fi

%% file: misc/data/non-sigstore-users-1.tex
\begin{table}[ht]
\centering
\small
\caption{
Non Sigstore Users: Tooling Issues causing a consideration of Sigstore and other Next-gen tools.
\JD{Move this to appendix.}
}
\normalsize
\small
\scriptsize
\label{tab:non_sigstore_reasons_for}
\begin{tabular}{p{0.15\linewidth} p{0.12\linewidth}p{0.60\linewidth}}
\toprule
\textbf{Tool} & \textbf{Subjects}  & \textbf{Topic} \\
\midrule
Internal tool      & P8, P11                         & \textit{\textbf{Tooling Issues}}\\
{} & {}  &(1) Integration with other tools/platforms\textbf{T} -- P8, P11 \\
{} & {}  &(2) Non unified tool implementation across teams \textbf{O} -- P8, P11 \\
{} & {}  &(3) Signer ID Management \textbf{T} -- P11 \\
{} & {}  &(4) Usage info/documentation unclear \textbf{T} -- P11 \\
{} & {}  &(5) Verification not built in with tool \textbf{T} -- P8 \\
{} & {}  &(6) Non transparency in signing infrastructure \textbf{T} -- P11 \\
{} & {}  &(7) Security of Signing Infrastructure \textbf{T} -- P11 \\
{} & {}  &(8) Engineers don’t understand the current tool \textbf{P} -- P11 \\
{} & {}  &\textit{\textbf{Considered (or Adopted) Fix}} \\
{} & {}  &(1) Organization building new tools with Next-gen features  -- P8, P11 \\
\midrule
Notary-V1       & P10                             & \textit{\textbf{Tooling Issues}}\\
{} & {}  &(1) Key Management\textbf{T} \\
{} & {}  &(2) Lack of an inbuilt signing policy \textbf{T}\\
{} & {}  &(3) Engineers don’t understand the current tool \textbf{P} -- P11 \\
{} & {}  &(4) Integration with other tools/platforms\textbf{T} \\
{} & {}  &\textit{\textbf{Considered (or Adopted) Fix}} \\
{} & {}  &(1) Organization has adopted Openpubkey for key management\\
{} & {}  &(1) Organization considering Sigstore, and Notary V2\\
\midrule
PGP           & P13                        & (1) \textit{\textbf{Tooling Issues}} \\
{} & {}  &(1) Key Management\textbf{T} \\
{} & {}  &(2)  Non transparency in signing infrastructure \textbf{T}\\
{} & {}  &(3) Security of Signing Infrastructure \textbf{T}\\
{} & {}  &\textit{\textbf{Considered (or Adopted) Fix}} \\
{} & {}  &(1) Use of Key discovery and delivery systems like TUF\footnote{The Update Framework\cite{tuf}}\\
\bottomrule
\end{tabular}
\end{table}